\documentclass{aa}  
\def\aap{AA}

\def\apjl{ApJL}

\def\apjs{ApJS}

\usepackage{graphicx}
\usepackage[titletoc]{appendix}
\usepackage{graphicx}
\usepackage{float}
\usepackage{amssymb}
\usepackage{amsfonts}
\usepackage{amsmath} 
\usepackage{color}
\usepackage{wrapfig}
\usepackage{hyperref}
\usepackage{breqn}
\usepackage{algorithm}
\usepackage{algpseudocode}
\usepackage{physics}

\usepackage{natbib}

\def\Ddt{{D_{\rm{\Delta t}}}}
\def\Ddteff{{D_{\Delta t}^{\rm{eff}}}}
\def\data{{\bf{d}}}
\def\qsub{{\bf{q}_{\rm{cdm}}}}
\def\msun{{M_{\odot}}}

\def\msub{{\bf{m}_{\rm{sub}}}}

\def\nuissance{{\bf{n}}}

\def\thetakX{{\boldsymbol{\theta_{k}^{X}}}}
\def\thetaktwoX{{\boldsymbol{\theta_{k+1}^{X}}}}

\usepackage{txfonts}
\begin{document} 
	
	\title{TDCOSMO III: Dark matter substructure meets dark energy - the effects of (sub)halos on strong-lensing measurements of $H_0$}
	
	
	\author{D. Gilman\inst{1}\thanks{gilmanda@ucla.edu},
		S. Birrer\inst{2},
		\and 
		T. Treu \inst{1}
	}
	\institute{Department of Physics and Astronomy, University of California,
		Los Angeles, CA 90095, USA
		\and
		Kavli Institute for Particle Astrophysics and Cosmology and Department of Physics, Stanford University, Stanford, CA 94305, USA
	}
	
	\date{\centering August 16th, 2020}
	
	\abstract
	{Time delay cosmography uses the arrival time delays between images in strong gravitational lenses to measure cosmological parameters, in particular the Hubble constant $H_0$. The lens models used in time delay cosmography omit dark matter subhalos and line-of-sight halos because their effects are assumed to be negligible. We explicitly quantify this assumption by analyzing mock lens systems that include full populations of dark matter subhalos and line-of-sight halos, applying the same modeling assumptions used in the literature to infer $H_0$. We base the mock lenses on six quadruply-imaged quasars that have delivered measurements of the Hubble constant, and quantify the additional uncertainties and/or bias on a lens-by-lens basis. We show that omitting dark substructure does not bias inferences of $H_0$. However, perturbations from substructure contribute an additional source of random uncertainty in the inferred value of $H_0$ that scales as the square root of the lensing volume divided by the longest time delay. This additional source of uncertainty, for which we provide a fitting function, ranges from $0.7 - 2.4\%$. It may need to be incorporated in the error budget as the precision of cosmographic inferences from single lenses improves, and sets a precision limit on inferences from single lenses.}
	
	
	\titlerunning{$H_0$ and dark matter substructure}
	\authorrunning{D. Gilman, et al.}
	\maketitle
	
	%
	\section{Introduction}
	
	Time delay cosmography with multiply-imaged quasars delivers measurements of the Hubble constant $H_0$ to percent-level precision \citep{Chen++19,Birrer++19a,Wong++19,Shajib++19a}, and substructure lensing studies use strong gravitational lenses to measure the halo mass function below $10^9 \msun$, thereby constraining the nature of dark matter \citep{DalalKochanek02,Nierenberg++14,Hezaveh++16b,Birrer++17,Vegetti++18,Hsueh++19,Gilman++20a, Gilman++20b}. To date, these two lines of research are pursued independently. Dark substructure is omitted from the lens models used in time delay cosmography, while substructure lensing analyses do not explicitly marginalize over the uncertainties of cosmological parameters. In this paper, we move to bridge the gap between these fields by quantifying how dark substructure affects the precision and accuracy of measurements of $H_0$ obtained from strong lenses.  
	
	The impetus to carry out this analysis now stems from a recent challenge to the standard $\Lambda$CDM cosmology. Strong-lensing measurements of $H_0$ in the local Universe presented by the H0LiCOW collaboration \citep{Suyu++17}, and from Type Ia supernova, carried out by the SH0ES collaboration \citep{Riess++19}, differ from the value of $H_0$ derived from the Cosmic Microwave Background \citep{Planck++18}. The  tension between the CMB data and six strongly-lensed quasars alone is 3.1$\sigma$, and exceeds $5 \sigma$ when lensing is combined with supernova data \citep{Wong++19,Riess19}. A resolution to this tension may require new physics. Before reaching this conclusion it is necessary to quantify potential sources of bias and measurement uncertainties relevant for time delay cosmography. 
	
	\citet{Oguri++07} pointed out that dark matter subhalos surrounding the main deflector in galaxy-scale strong lenses alter the arrival times between lensed images. The arrival times provide a direct, one-step method to measure $H_0$, so it is conceivable that perturbations from substructure to these data could bias, or contribute to the error budget, of cosmographic analyses. \citet{KeetonMoustakas09} \citep[see also][]{Congdon++10,Cyr-Racine++16} carried out the first comprehensive analysis of time delay perturbations from substructure, and showed that populations of dark subhalos impart arrival time perturbations on the scale of a few hours. This level of perturbation lies safely below current measurement precision, which  measures image arrival times down to a few-days precision \citep[e.g.][]{Bonvin++18,Courbin++18b,Millon++20}. On these grounds, substructure is assumed to be negligible and it is omitted from the strong lens models used for cosmography. 
	
	However, as the precision of time delay cosmography improves it is timely to revisit this assumption and quantify the potential impact of substructure both in terms of accuracy and precision. First, \citet{KeetonMoustakas09} did not account for field halos along the line of sight to strong lenses, which can significantly outnumber subhalos. Because the number of line-of-sight halos depends on the lens and source redshifts, the amplitude of time delay perturbations should evolve with redshift, as should any resulting uncertainties and/or bias in the $H_0$ value inferred from the time delays. Second, it is possible that perturbations from dark matter halos on lensing observables propagate to $H_0$ inferences through indirect channels, not directly through the time delays. The collective lensing effects of a population of halos could be absorbed by the global mass model of the lens system, or the \textit{macromodel}, in which case substructure would impart stochastic perturbations to the macromodel. 
	
	In this work, we quantify these effects by applying the same modeling assumptions used in cosmographic analyses to mock lenses that include full populations of subhalos and line-of-sight halos. Using the image positions and the surface brightness of lensed arcs to constrain the macromodel, we infer $H_0$ from the mock lenses with lens models that do not include substructure, as is currently done in the literature. Comparing the inferred $H_0$ with the `true' value of $H_0$ used to create the mocks, we quantify to what degree dark substructure biases or increases the uncertainties reported in cosmographic inferences, had it been included in the lens model. The mock lenses are based on six real quadruply-imaged quasars systems modeled by the H0LiCOW, STRIDES, and SHARP teams (from now on, we refer to this collaboration as TDCOSMO) that have delivered measurements of $H_0$, allowing us to estimate the effects of substructure for each individual system. By construction, our simulations isolate the role of substructure from other sources of uncertainty, such as the mass sheet transformation \citep{SchneiderSluse13} and systematic uncertainties associated with the lens macromodel \citep[e.g.][]{Kochanek19}. 
	
	A necessary component of our analysis is the creation of hundreds of mock lenses with full line-of-sight and subhalo populations. The dark matter halos in the mock systems simultaneously perturb the arrival times and the flux ratios. In addition, halos occasionally produce gravitational imaging residuals in extended images that lead to the discovery of individual dark matter halos in strong lens systems, a technique known as gravitational imaging \citep{Koopmans++05,Vegetti++10,Hezaveh++16b,Vegetti++18}. Based on our simulations, we comment on results relevant for gravitational imaging, and in particular recent suggestions to measure the convergence power spectrum of dark matter halos \citep[e.g.][]{Hezaveh++16b,Cyr-Racine++16}. As the focus of this work is time delay cosmography, we relegate discussion of these topics to Appendices \ref{app:imagingresiduals} and \ref{app:apptdelayfr}. 
	
	This paper is organized as follows: Section \ref{sec:methodology} describes our approach for estimating $H_0$ uncertainties from simulations that include substructure in mock lenses. Section \ref{sec:setup} describes the specific modeling choices implemented in the simulations. Section \ref{sec:results} describes the results of the analysis, which we summarize concisely in Section \ref{sec:summary}. We make concluding remarks in Section \ref{sec:conclusions}. 
	
	We perform gravitational lensing computations with {\tt{lenstronomy}}\footnote{https://github.com/sibirrer/lenstronomy} \citep{BirrerAmara18}. Computations that involve the halo mass function and the matter power spectrum are executed with {\tt{colossus}} \citep{Diemer17}. We assume a flat $\Lambda$CDM cosmology using the parameters from WMAP9 \citep{WMAP9cosmo} ($\Omega_m = 0.28, \sigma_8 = 0.82$), and set $h=0.733$ (see Section \ref{sec:methodology}). The conclusions we present are independent of the choice of cosmological parameters. 
	
	\section{Cosmographic inference on mock lenses that include substructure}
	\label{sec:methodology}
	This section describes our method to estimate the effect of dark substructure on cosmographic inferences. A brief outline is as follows: First, we create mock datasets with substructure included in the lens models. Next, we analyze the data as would be done in a modern cosmographic inference, ignoring substructure in the model. Finally, we investigate how the omission of dark substructure affects cosmographic inferences through comparison with a reference dataset that contains no substructure. 
	
	We begin by introducing the time delay distance $\Ddt$, and describe the observables from strong lens systems used to constrain it, in Section \ref{ssec:data}. In Section \ref{ssec:outline}, we discuss the task of inferring $H_0$ in the presence of full populations of dark matter halos in a Bayesian context. Section \ref{ssec:creatingdata} describes how we create mock datasets with substructure included the lens model. Section \ref{ssec:fitting} details our procedure for using these mock datasets to quantify any bias or additional uncertainties associated with the presence of substructure in strong lenses. 
	
	\subsection{The time delay distance and strong lensing observables}
	\label{ssec:data}
	Strong lensing constrains the Hubble constant through a measurement of an absolute distance scale $\Ddt$, referred to as the \textit{time delay distance} $\Ddt = \left(1+z_d\right) \frac{D_{\rm{d}} D_{\rm{s}}}{D_{\rm{ds}}}$. The factors $D_{\rm{d}}$, $D_{\rm{s}}$, $D_{\rm{d,s}}$ are angular diameter distances to the deflector, the source, and from the deflector to the source\footnote{Through the text $D_{\rm{i,j}}$ stands for an angular diameter distance from the $i$th lens plane to the $j$th. A subscript $s$ stands for the source plane.}, and $z_d$ is the deflector redshift. The combination of cosmological distances entering $\Ddt$ depends on cosmology, with a particularly strong dependence on $H_0$, with $\Ddt \propto \frac{1}{H_0}$. 
	
	The data vector $\data = \left(\boldsymbol{\theta}_{\rm{img}}, \boldsymbol{t}, \boldsymbol{s}\right)$  that constrains quantities such as $\Ddt$ includes the four lensed quasar image positions $\boldsymbol{\theta}_{\rm{img}} = \left(\boldsymbol{\theta_A}, \boldsymbol{\theta}_B, \boldsymbol{\theta}_C, \boldsymbol{\theta}_D\right)$, the time delays between the images $\boldsymbol{t} = \left(t_{\rm{12}}, t_{\rm{13}}, t_{\rm{14}}\right)$, and the observed surface brightness $\boldsymbol{s}\left( \boldsymbol{\theta}\right)$ at angular position $\boldsymbol{\theta}$ in the image plane. 
	
	All matter in the lens system, dark or baryonic, affects the time delays, image positions, and the arc surface brightness. Two equations encode the dependence of these observables on the deflection angles $\boldsymbol{\alpha}$, and on angular diameter distances and cosmology. First, the ray tracing equation describes the propagation of light rays through lines of sight populated by dark matter halos \citep{BlandfordNarayan86}
	\begin{equation}
	\label{eqn:raytracing}
	\boldsymbol{\theta_K}\left(\boldsymbol{\theta}\right) = \boldsymbol{\theta} - \frac{1}{D_{\rm{s}}} \sum_{k=1}^{K-1} D_{\rm{k,s}}{\boldsymbol{\alpha_{\rm{k}}}} \left(D_{\rm{k}} \boldsymbol{\theta_{\rm{k}}}\right).
	\end{equation} 
	This is a recursive relation for angular coordinates on the $k$th lens plane, given a coordinate on the sky $\boldsymbol{\theta}$, the deflection angle $\boldsymbol{\alpha}$, and the angular diameter distance from the $k$th lens plane to the source plane, $D_{\rm{k,s}}$. Equation \ref{eqn:raytracing} gives the surface brightness in the image plane $s\left(\boldsymbol{\theta}\right)$ by ray-tracing through the lens model to the source plane, and the multiple images of the background quasar appear at the coordinates $\boldsymbol{\theta}$ that each map to the same position on the source plane. 
	
	The time delay between images located at coordinates $\boldsymbol{\theta_{A}}$ and $\boldsymbol{\theta_{B}}$ is
	\begin{equation}
	\label{eqn:tabmulti}
	t_{\rm{AB}} = \frac{D_{\Delta t}^{\rm{eff}}}{c} \left(\phi^{\rm{eff}} \left(\boldsymbol{\theta_{A}}\right) - \phi^{\rm{eff}} \left(\boldsymbol{\theta_{B}}\right)\right).
	\end{equation}
	where the effective Fermat potential is given by 
	\begin{equation}
	\label{eqn:fermatpoteff}
	\begin{split}
	\phi^{\rm{eff}} \left(\boldsymbol{\theta_{X}}\right) \equiv & \sum_{k=1}^{K-1} \frac{1+z_k}{1+z_{\rm{d}}} \frac{D_k D_{k+1} D_{\rm{ds}}}{D_{\rm{d}} D_{\rm{s}} D_{k,k+1}} \\
	& \times \left[\frac{\left(\thetakX - \thetaktwoX\right)^2}{2} - \xi_{k,k+1} \Psi\left(\thetakX\right) \right],
	\end{split}
	\end{equation}
	where $\Psi\left(\boldsymbol{\theta}\right)$ is the projected gravitational potential at position $\boldsymbol{\theta}$, $\xi_{i,j} \equiv \frac{D_{i,j} D_s}{D_{i} D_{i,s}}$, and where we use shorthand notation $\boldsymbol{\theta_K}\left( \boldsymbol{\theta_{X}} \right) \equiv \boldsymbol{\theta_{K}^{X}}$ for Equation \ref{eqn:raytracing} evaluated at a specific angular position $\boldsymbol{\theta_{X}}$. The subscript $d$ refers to the main deflector lens plane.
	
	Assuming a cosmology, there is a unique distance scale $\Ddteff$,  given a value of $H_0$. Following the standard approach \citep[e.g.][]{Wong++19,Shajib++19b}, we can define an effective time delay distance $\Ddteff$ that normalizes the measured time delays and the model-predicted Fermat potential, and compute $H_0$ from this effective distance scale. 
	
	A typical cosmographic analysis includes nuisance parameters $\nuissance$, such as the lens mass profile, the lens light profile, the source light profile, and other factors that can affect the inference such as the microlensing time delay \citep{TieKochanek18}, and the external convergence associated with large-scale structure near the lens and along the line of sight \citep{SchneiderSluse13,Birrer++16}. Obtaining a $\Ddt$ posterior from a strong lens system requires marginalizing the likelihood function over these nuisance parameters
	\begin{equation}
	\label{eqn:likelihood1}
	p\left(\data | \Ddt \right) = \int \mathcal{L}\left(\data | \Ddt, \nuissance \right) p\left(\nuissance \right) d \nuissance.
	\end{equation}
	The evaluation of Equation \ref{eqn:likelihood1} involves simultaneously reconstructing the lensed background source and the lensed image. 
	
	\subsection{Directly computing $H_0$ posteriors with substructure}
	\label{ssec:outline}
	
	To incorporate substructure, we describe dark field halos and subhalos with population-level statistical properties specified by a set of hyper-parameters $\qsub$. These hyper-parameters describe, for example, the form of the field halo and subhalo mass functions, the average concentration-mass relation of halos, the probability distribution for their spatial distribution, etc., assuming Cold Dark Matter (CDM). Typically, cosmographic analyses ignore $\qsub$ altogether when measuring $\Ddt$, assuming $p\left(\Ddt | \data, \qsub \right) \approx  p\left(\Ddt | \data \right)$. In other words, the assumption in the literature is that inferences on $H_0$ do not depend on the nature of small-scale dark matter structure in the lens system, or at least that other factors in the analyses, such as the nuisance parameters $\nuissance$, dominate the $\Ddt$ error budget such that dark substructure can be safely ignored. 
	
	The direct route towards testing this assumption involves a daunting calculation of a likelihood function. We can account for dark substructure in Equation \ref{eqn:likelihood1} by generating realizations of dark matter halos specified by parameters $\msub$ from the model specified by hyper-parameters $\qsub$. The likelihood that accounts for $\qsub$ includes an additional integral over all possible configurations $\msub$
	\begin{equation}
	\label{eqn:likelihood2}
	\begin{split}
	\mathcal{L}\left(\data | \Ddt, \qsub\right) =& \int p\left(\data | \Ddt, \msub, \nuissance \right) \\
	& \times p\left(\msub | \qsub\right) p\left(\nuissance \right) d \msub d \nuissance
	\end{split}
	\end{equation}
	
	Incorporating substructure by directly evaluating the integral over $\msub$ is not feasible because Equation \ref{eqn:likelihood2} is computationally intractable. The problem stems mainly from integrating $p\left(\data | \Ddt, \msub, \nuissance \right)$ over $\nuissance$, because $\msub$ includes potentially tens of thousands of halos whose deflection angles need to be continuously re-evaluated while performing the integral over $\nuissance$. 
	
	\subsection{Forward modeling substructure perturbations on lensing data}
	\label{ssec:creatingdata}
	To circumvent the direct evaluation of the intractable integral in Equation \ref{eqn:likelihood2}, we instead create mock datasets computed with substructure in the lens model using a a known `true' value of $H_0$. We then apply same assumptions used in the literature (ignore substructure) to analyze these systems with the goal of recovering the `true' $H_0$ value. The degree to which we fail to recover the `true' $H_0$, and the degree to which the uncertainties increase relative to a control sample with no substructure in the lens system, allowing us to to quantify any bias or increased uncertainties associated with substructure. 
	
	The mock lenses we generate are analogs of the lens systems RXJ1131-1231, PG1115+080, HE0435-1223, B1608+656, WFI2033-4723, and DESJ0408+5354, each of which has been used to measure $H_0$ (see the discussion in Section \ref{ssec:individualdis} for details). Through the procedure outlined in the remainder of this section, we determine on a case-by-case basis to what degree omitting substructure from the lens model results in under-estimated uncertainties and/or bias in the inferred value of $H_0$.
	
	To begin, we describe two sets of mock data created for each lens system: the \textit{control data} and the \textit{substructure-perturbed data}:
	
	\begin{itemize}
		\item \textit{The control data}: The control data includes no substructure, and the same parameterization for the mass model is used to create the data as is used to fit the resulting mock lens. The uncertainty in $H_0$ inferred from this dataset is determined entirely by the measurement precision of the time delays, and the degree to which the lensed images and extended arc constrain the mass profile. We tune the signal to noise in the mock data until the $H_0$ inference we obtain from this dataset has the same error budget from time delays and lens modeling as those quoted in Table 3 of \citet{Wong++19}, or the uncertainties reported by \citet{Shajib++19b} in the case of DESJ0408. 
		
		\item \textit{The substructure-perturbed data}: We first generate a realization of dark matter field halos and subhalos. Next, starting from the lens model used to create the control dataset, we solve for a new set of macromodel parameters that map the observed image positions to the same coordinate on the source plane through Equation \ref{eqn:raytracing}. The new macromodel is initialized from the macromodel used to create the control data, so the new macromodel is nearby in parameter space to the one used to create the control dataset. We place the same extended background source used to create the control data at the new source position obtained from adjusting the macromodel, and proceed to compute the time delays and the surface brightness of the extended images. The mock lens obtained through this process is indistinguishable by eye from the control data, but the time delays and lensed arc include the subtle lensing signatures of dark matter substructure. 
	\end{itemize}
	
	\noindent Both the control and substructure-perturbed models are created using the same value of $H_0 = 73.3 \ \rm{km} \ \rm{s^{-1}} \rm{Mpc}$, the value quoted by \citet{Wong++19}, and we apply the same Gaussian PSF with a FWHM of 50 milli-arcseconds to the simulated lenses, comparable to the imaging resolution from the Hubble Space Telescope. We stress that our results are independent of the specific choice of $H_0$.
	
	\subsection{Estimating $H_0$ uncertainties from un-modeled dark substructure}
	\label{ssec:fitting}
	We fit the mock datasets using smoothly-parameterized mass profiles for the main deflector mass profile while ignoring substructure in the lens system. The model applied to the control data is the same as the model used to create the data, so we should fit the mock data down to the noise; this is a perfect model scenario. We then apply the same modeling assumptions to the substructure-perturbed datasets, fitting them with the same smoothly-parameterized mass profile. To isolate the role of substructure, we assume perfectly-measured image positions and do not add astrometric uncertainties to the lensed point sources. The flux ratios are not used to constrain the lens model, although we can compute them jointly with the time delays (see Appendix \ref{app:apptdelayfr}). We perform the lens modeling on 200 mock datasets, each of which corresponds to one realization of dark matter halos. This gives 200 corresponding posterior distributions of $H_0$, which we combine by adding them together. 
	
	In our procedure, mimicking the actual measurements,  we include additional model complexity in the mock datasets, rather than in the model applied to the data. The requirement that must be met in order to safely make this approximation is that each of the substructure-perturbed datasets wield the same constraining power over $H_0$. We take two measures to ensure this is the case: first, because the mock `measured' time delays change slightly from realization to realization in the substructure-perturbed datasets, a fixed uncertainty in units of days translates to different percent precision in $H_0$. To keep the same constraining power of the time delays the same for each individual realization, we operate with \textit{relative} uncertainties on the time delays, using the same uncertainties as those measured from the real lens systems studied in the literature. 
	
	Second, by re-weighting posterior samples we ensure that we operate at fixed uncertainty in the logarithmic slope of the main deflector mass profile $\gamma$, and the Einstein radius $R_{\rm{Ein}}$. Operating at fixed uncertainty in $\gamma$ and $R_{\rm{Ein}}$ serves two purposes: first, power-law mass models fit to the lens system HE0435-1223 (to use a specific example) have logarithmic slopes $\gamma = 1.9 \pm 0.05$ \citep{Chen++19}. Substructure realizations that lead to global mass profiles with profile slopes substantially different than this are therefore not analogs of the lens in question. Quantitatively, the realizations that lead to this mismatch between the model and the data do not contribute to the integral in Equation \ref{eqn:likelihood2}, because the likelihood associated with these realizations would be extremely low. This argument applies to both $\gamma$ and $R_{\rm{Ein}}$. The second reason for operating at fixed $\gamma$ and $R_{\rm{Ein}}$ is that both of these parameters are, in principle, constrained by the stellar kinematics of the main deflector. Enforcing equality between the $\gamma$ and $R_{\rm{Ein}}$ posteriors ensures that the modeled-predicted velocity dispersion matches the velocity dispersion that would have been `measured' from the mock datasets. By incorporating information that could, in principle, be constrained by kinematic data, we isolate the role of substructure in the lens system. 
	
	After this series of steps, we are ready to make comparisons between the $H_0$ inference from the control data and the inferences from the substructure-perturbed data. If substructure is irrelevant for time delay cosmography, then we expect the $H_0$ posteriors obtained from the control data and the substructure-perturbed dataset to be identical. The degree to which the $H_0$ posteriors differ allows us to quantify how the reported values of $H_0$ from the TDCOSMO analyses would have been affected, had substructure been included in the lens model. Compared to the control data, the $H_0$ inferences from the substructure-perturbed data will be inflated by some amount $\sigma_{\rm{sub}}$, independent of the other sources of uncertainty  $\sigma_{\rm{reported}}$. The total uncertainty is therefore
	\begin{equation}
	\sigma_{\rm{total}}^2 = \sigma_{\rm{reported}}^2 + \sigma_{\rm{sub}}^2.
	\end{equation}
	A bias in $H_0$ will manifest as different medians between the control and substructure-perturbed posteriors. 
	
	\section{Simulation Setup}
	\label{sec:setup}
	\begin{figure*}
		\includegraphics[clip,trim=0.5cm 3.5cm 0.3cm
		5cm,width=.95\textwidth,keepaspectratio]{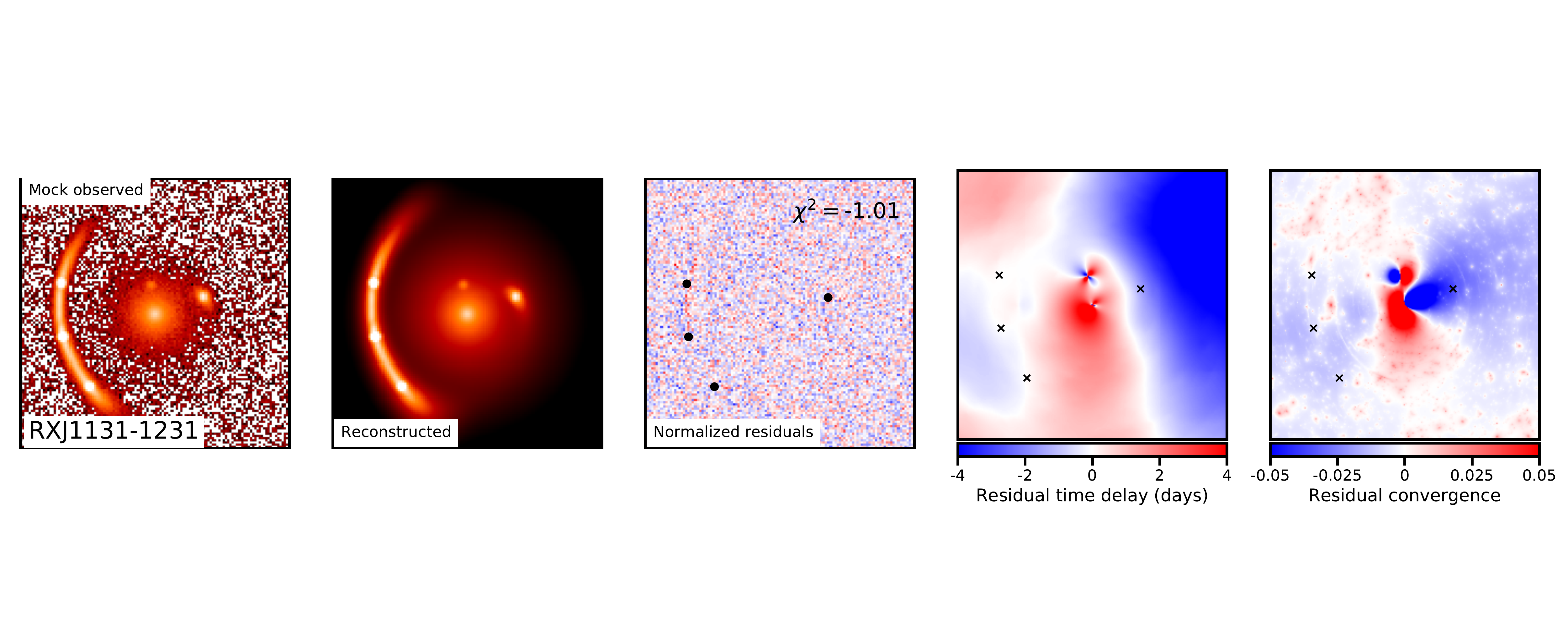}
		\includegraphics[clip,trim=0.5cm 3.5cm 0.3cm
		5cm,width=.95\textwidth,keepaspectratio]{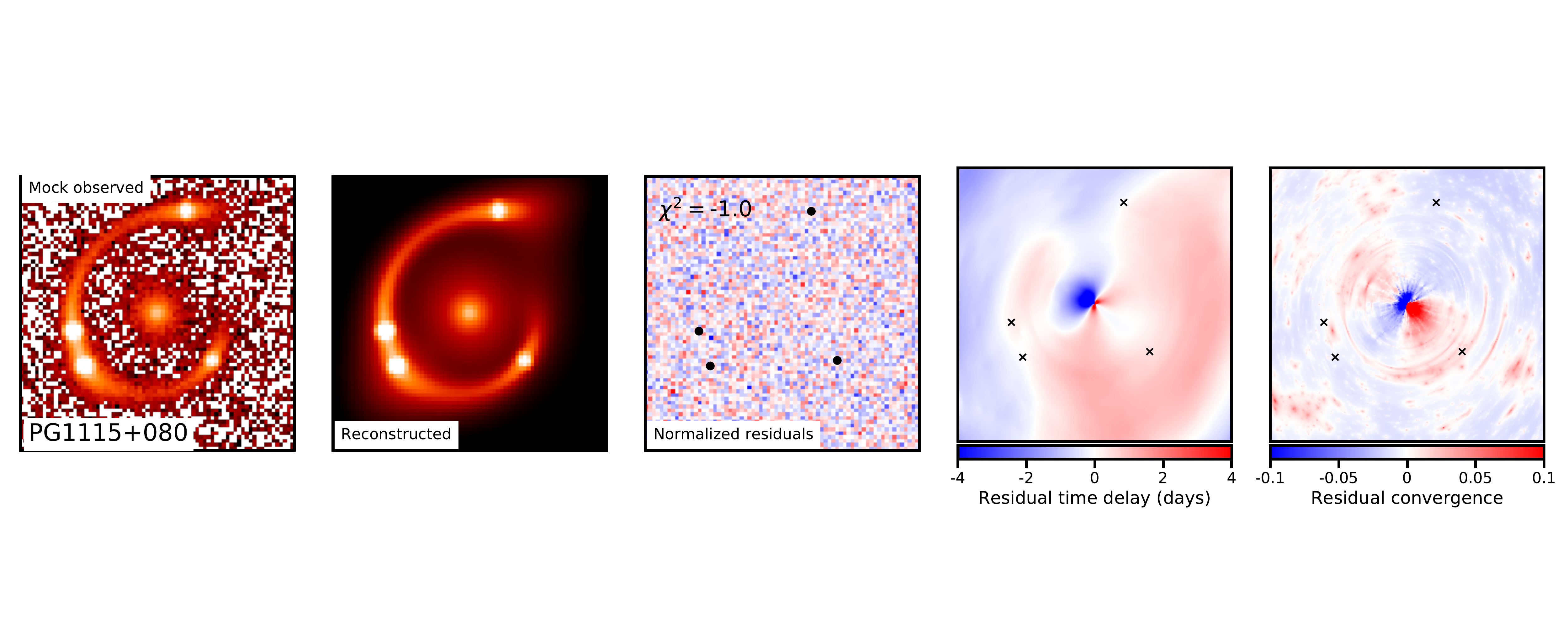}
		\includegraphics[clip,trim=0.5cm 3.5cm 0.3cm
		5cm,width=.95\textwidth,keepaspectratio]{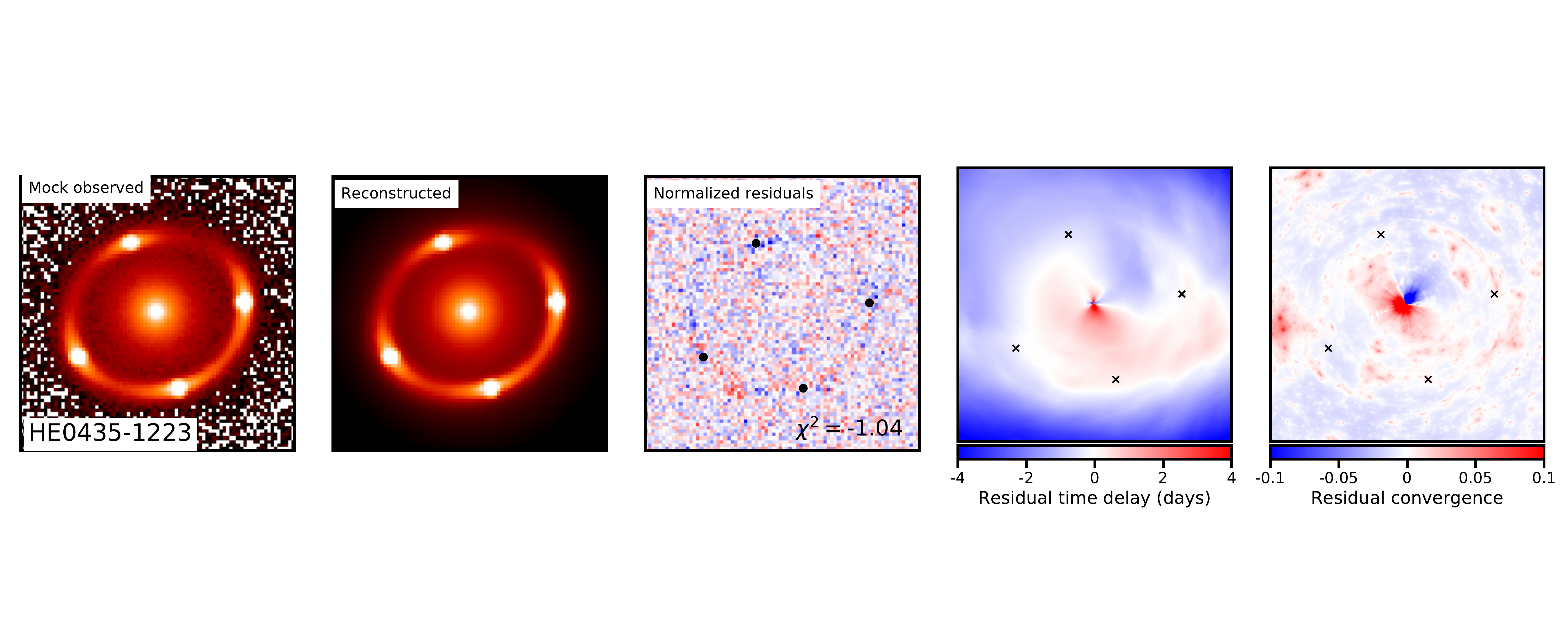}
		\caption{\label{fig:lensmaps1} The panels show the mock observed lens that includes substructure in the lens model (far left), the reconstructed lensed image (second from left), the normalized imaging residuals (center), the residual time delay surface (second from right), and the residual convergence (far right). The residual time delay surface is the `true' time delay surface from the mock lens minus the model-predicted time delay surface from the lens model fit to the data. Similarly, the residual convergence is the full multi-plane convergence (Equation \ref{eqn:multiplaneconv}) minus the convergence from the model fit to the mock data. The lensed quasar image positions are marked as points (crosses) in the center (right) panels. Each row corresponds to one substructure-perturbed dataset for the lens systems based on RXJ1131-1231, PG1115+080, and HE0435-1223.}
	\end{figure*}
	
	\begin{figure*}
		\includegraphics[clip,trim=0.5cm 3.5cm 0.3cm
		5cm,width=.95\textwidth,keepaspectratio]{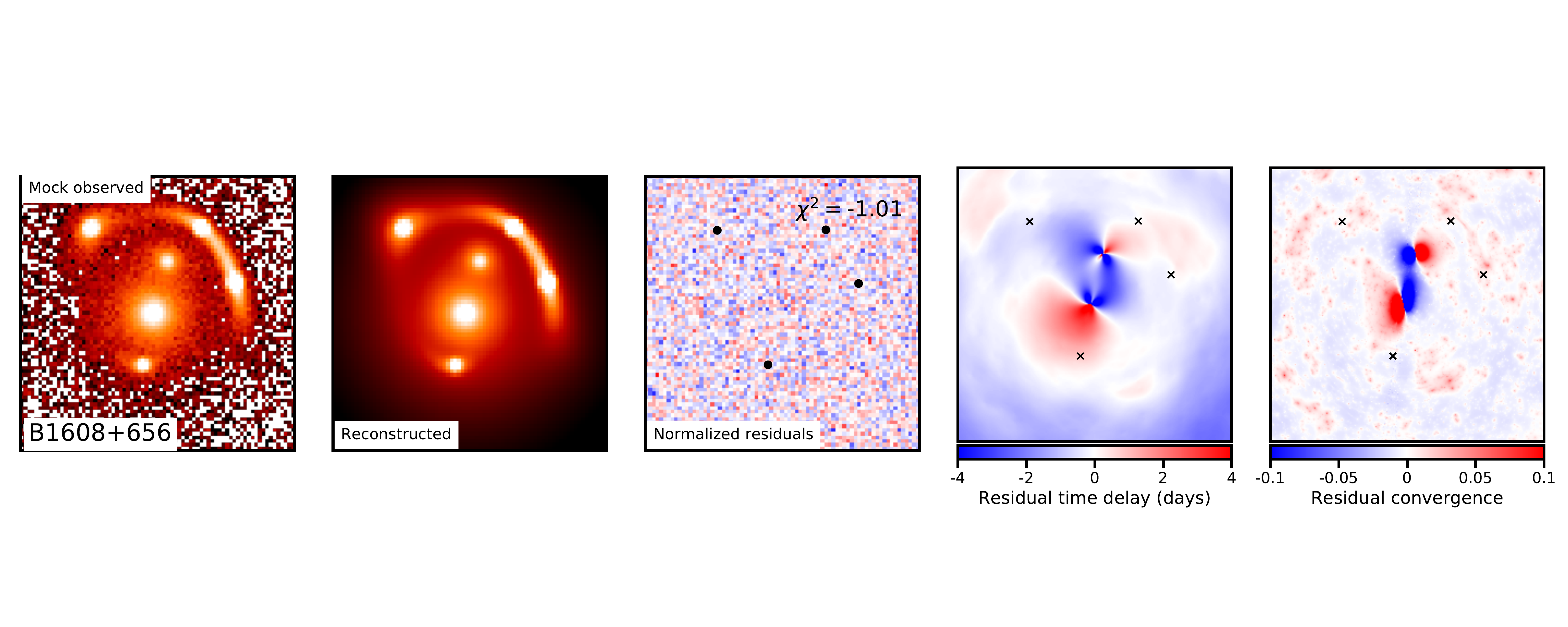}
		\includegraphics[clip,trim=0.5cm 3.5cm 0.3cm
		5cm,width=.95\textwidth,keepaspectratio]{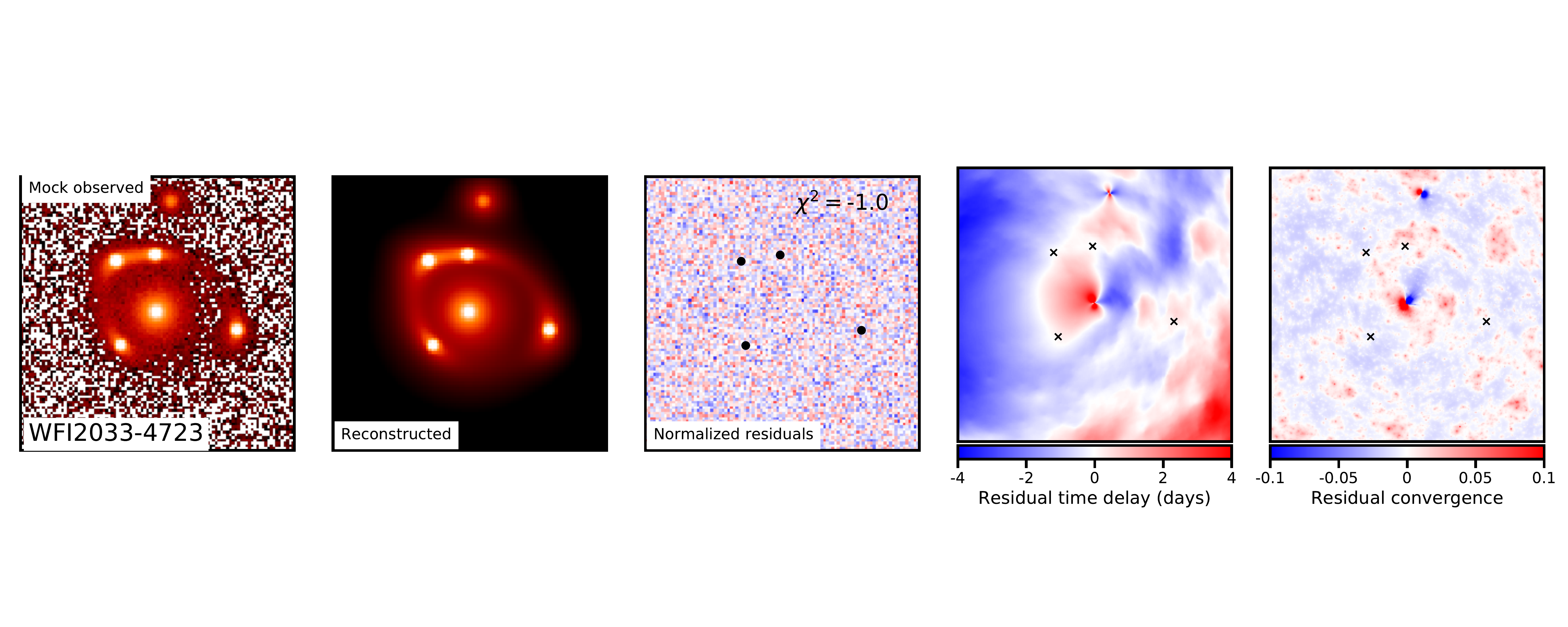}
		\includegraphics[clip,trim=0.5cm 3.5cm 0.3cm
		5cm,width=.95\textwidth,keepaspectratio]{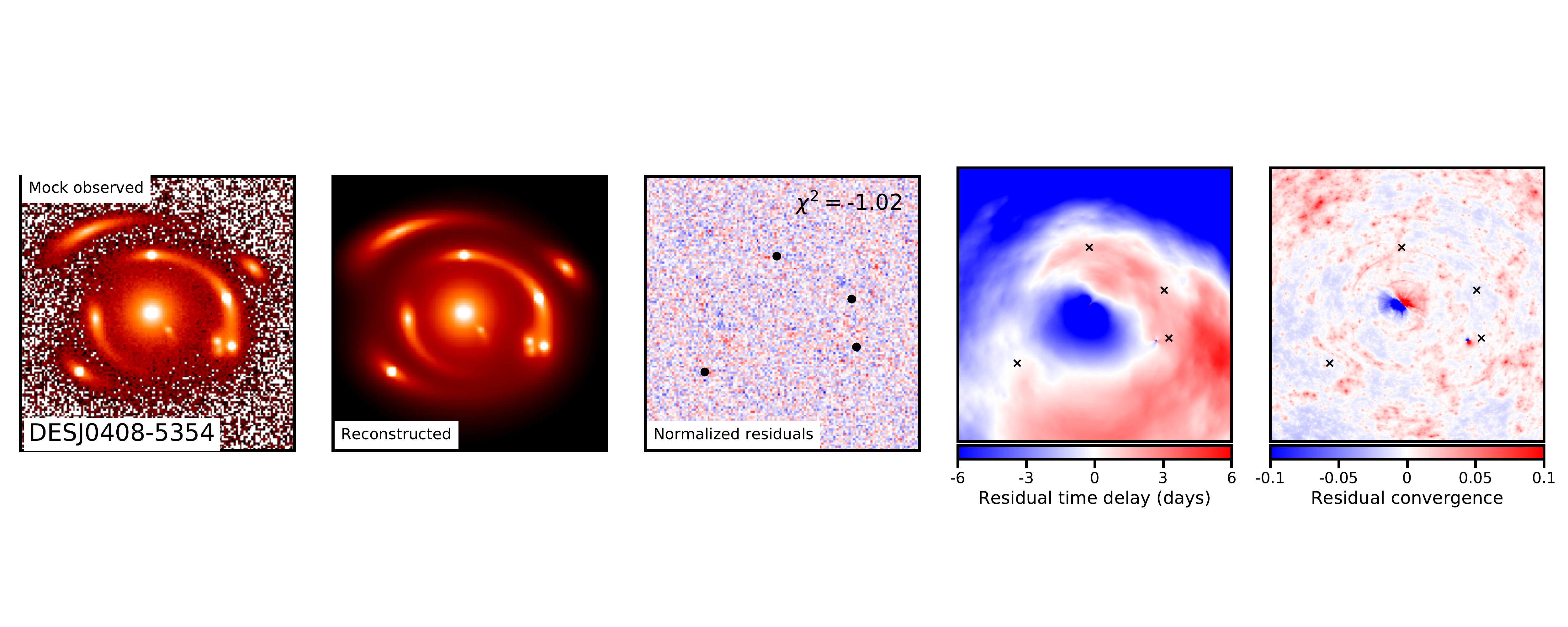}
		\caption{\label{fig:lensmaps2} The figure is the same as Figure \ref{fig:lensmaps1}, but shows examples of the mock observed and reconstructed lens (left), the imaging residuals (center), the residual time delay surface (second from right), and the residual convergence (far right) maps for the lens systems B1608+656, WFI2033-4723, and DESJ0408-5354.}
	\end{figure*}
	
	The procedure outlined in the previous section requires a prescription for rendering populations of dark matter halos, and models for the lens mass distribution, lens light profile, and source light profile. We describe the parameterizations and modeling choices for the subhalo and line-of-sight halo populations in Sections \ref{ssec:rendering}. Section \ref{ssec:symmetry} describes how subhalos and line-of-sight halos are generated symmetrically around the light rays throughout the lensing volume, and the addition of negative convergence profiles to correct for the additional mass added in substructure. Sections \ref{ssec:massmodels} and \ref{ssec:sourcemodels} describe the models we implement for the main deflector mass profile and the background source light profile, respectively. We reiterate that the lens models we use omit complexity in the form of external convergence or more complicated mass profiles in order to isolate the perturbative effects of substructure. 
	
	\subsection{The CDM subhalo and field halo populations}
	\label{ssec:rendering}
	We use the same parameterizations for the CDM mass function as those implemented by \citep{Gilman++20a,Gilman++20b}; we briefly review the essential features in the following paragraphs.  
	
	Both subhalos and field halos along the line of sight, are modeled as truncated NFW profiles \citep{Navarro++96,Baltz++09} defined with a density contrast of 200 with respect to the critical density at $z=0$. We use the mass-concentration relation presented by \citet{DiemerJoyce20}. The truncation radius for subhalos depends on the mass of the subhalo, and the position inside the host halo, while field halos are truncated at $r_{50}$. The choice of $r_{50}$ for field halos is motivated by the splashback radius, a physical halo boundary that is typically larger than, but the same order of magnitude, as the virial radius \citep{Diemer20}. The truncation of field halos at radii $> r_{200}$ keeps the total mass in line-of-sight halos finite with a negligible impact on their lensing properties.  
	
	We draw field halo masses from a Sheth-Tormen mass function \citep{ST99} with the addition of the two-halo term $\xi_{\rm{2halo}}\left(M_{\rm{halo}}, z\right)$  
	\begin{equation}
	\label{eqn:losmfunc}
	\frac{d^2N_{\rm{los}}}{dm  dV} = \big(1+ \xi_{\rm{2halo}}\left(M_{\rm{host}}, z\right)\big) \frac{d^2N}{dm  dV} \big \vert_{\rm{ShethTormen}}.
	\end{equation}
	The two-halo term boosts the amount of structure near the host halo, which has a mass $M_{\rm{host}}$. We render subhalos from a power-law mass function defined in projection
	\begin{equation}
	\label{eqn:subhalomfunc}
	\frac{d^2 N_{\rm{sub}}}{dm dA} =  \frac{\Sigma_{\rm{sub}}}{m_0} \left(\frac{m}{m_0}\right)^{\alpha} \mathcal{F} \left(M_{\rm{host}}, z\right)
	\end{equation}
	with logarithmic slope $\alpha = -1.9$ \citep{Springel++08} and a pivot mass $m_0 = 10^{8} \msun$. The function $\mathcal{F} \left(M_{\rm{host}}, z\right)$ encodes the evolution of the projected number density of subhalos with host halo mass and redshift \citep[see][]{Gilman++20a}, and is calibrated using the semi-analytic modeling code {\tt{galacticus}} \citep{Benson12}. It is given by $\log_{10} \left(\mathcal{F}\right) = k_1 \log_{10}\left(\frac{M_{\rm{host}}}{10^{13} M_{\odot}}\right) + k_2 \log_{10} \left(1+z\right)$, with $k_1 = 0.88$ and $k_2 = 1.7$. We assume a host halo mass of $10^{13.3} \msun$ for each mock lens, based on the population mean halo mass for strong lensing galaxies inferred by \citet{Lagattuta++10}. We use a normalization $\Sigma_{\rm{sub}} = 0.025 \ \rm{kpc}^{-2}$ based on the measurements from \citep{Gilman++20a}. These modeling choices are intended to be representative of the subhalo populations around massive elliptical galaxies. 
	
	We render both halos and subhalos in the mass range $10^6 - 10^9 \msun$. We  have carried out tests with a lower minimum mass threshold, and verify that our results do not depend on this choice. We assume halos more massive than $10^9$ would host a visible galaxy, and would therefore be modeled explicitly in cosmographic analyses. The choice of $10^9$ solar masses is conservative in the sense that a larger mass would increase the uncertainties we attribute to un-detected substructure. The rendering volume is a double-cone geometry with an opening angle of 15 times the Einstein radius of each lens. We have tested larger opening angles, and verify that our results are unchanged. The total number of halos and subhalos included in lens models ranges from $50,000 - 200,000$, depending on the lens and source redshifts, and the Einstein radius. 
	
	\subsection{Lens cone symmetry and convergence corrections}
	\label{ssec:symmetry}
	At each lens plane along the line of sight\footnote{Lens planes are placed uniformly in redshift with a spacing of 0.02.}, we generate halos uniformly distributed in two dimensions around a coordinate center that tracks the path of a lensed light ray backwards from the observer towards the lens mass centroid. Behind the main deflector, the rendering area tracks the light as it is deflected by background galaxies, as is the case in HE0435, WFI2033, and DESJ0408. Without shifting the rendering area to track lensed light rays, substructure is not distributed symmetrically around the path of the light, and the asymmetric mass distribution biases the inferred value of $H_0$. The shifting of substructure along a lensed light path is particularly important for lens systems with large satellites and/or line-of-sight galaxies, as the source coordinate in these systems can be offset from the lens center by as much as $1 \ \rm{arcsec}$, comparable to a typical lens Einstein radius. 
	
	We add negative convergence at each lens plane to subtract the mean mass rendered in (sub)halos, ensuring that the average dark matter density in the rendering volume is equal to the mean dark matter density of the Universe. For line-of-sight halos, the correction is in the form of a uniform (negative) convergence sheet. The correction is more complicated for subhalos, as they are spatially distributed following a cored NFW profile. The cored spatial distribution, with a central core radius set to half the scale radius of the host, is intended to mimic the tidal disruption of subhalos that pass close to the central galaxy. The net convergence profile that needs to be subtracted for the subhalos is formally given by a convolution of the subhalo density profile with their spatial distribution function, but we are able to approximate this profile with a simple cored NFW profile normalized such that the residual convergence from substructure near the Einstein radius is zero on average. 
	
	Omitting the mass sheet corrections trivially biases $H_0$ by re-scaling each time delay through the addition of a net positive convergence in the form of dark matter halos. This effect is similar to the external convergence that is accounted for in cosmographic inferences, which affects the inferred time delay distance as
	\begin{equation}
	\Ddt = \frac{\Ddt_{\rm{model}}}{1 - \kappa_{\rm{ext}}}.
	\end{equation}
	While the ensemble of substructure-perturbed datasets have a residual convergence of zero on average, on a realization-by-realization basis the net residual convergence from substructure fluctuates around the mean dark matter density in the Universe. This introduces a mass-sheet transformation from random populations of dark matter halos that is distinct from the external convergence $\kappa_{\rm{ext}}$, which comes from large-scale structure in the lensing volume. The fluctuations of the substructure convergence cause fluctuations in the model-predicted time delays, and therefore contribute to the $H_0$ error budget. 
	
	\subsection{The main deflector mass and light profiles}
	\label{ssec:massmodels}
	Since our goal is to isolate the effects of substructure, we use the same mass and light profiles to create the control and substructure-perturbed datasets as we use in the lens models fit to these data. The main deflector is modeled as a power-law ellipsoid with logarithmic mass profile slope $\gamma$, plus external shear $\gamma_{\rm{ext}}$. To keep the analogs as similar to the real lenses as possible, we choose a logarithmic profile slope $\gamma$ consistent with the mean value inferred in the lens modeling by TDCOSMO. While the degree to which the macromodel absorbs the effects of substructure depends in principle on how it is parameterized, we do not expect the specific choice of a power law ellipsoid to significantly affect our results. Substructure imparts lensing effects on much smaller scales that those that can be subsumed into most reasonable parameterizations of the macromodel.\footnote{The exception to this is an effective external shear component from substructure that is absorbed by the external shear applied across main lens plane.}
	
	All of the lens systems we consider have either a nearby satellite galaxy in the main lens plane, a massive galaxy along the line of sight near the main deflector (in projection), or both. We model these objects as singular isothermal spheres, using Einstein radii and positions inferred by the lens modeling from the TDCOSMO teams. We add an elliptical S\'{e}rsic light profile for the main deflector light, and circular S\'{e}rsic light profiles for the satellite or nearby galaxies. The parameters describing the macromodels used to to create the mock lenses are listed in Table \ref{tab:baselinemodels}.
	
	\subsection{Background source light profiles}
	\label{ssec:sourcemodels}
	The mock datasets are all created using an elliptical S\'{e}rsic model for the background source. The exact parameters describing these profiles are unique to each lens, and are chosen such that the lensed arc in each mock resembles the extended features in the real lens it is based on.  
	
	Modeling lenses for cosmography involves a simultaneous reconstruction of the background source light and the lensed image. Using an elliptical S\'{e}rsic profile to model the background source (the same model as is used to create the mock lens), we detect significant residuals in the reconstructed lensed images that result from substructure in the lens system. Following the approach taken in the literature, when we detect these residuals we add additional complexity to the background source. We add complexity to the background source model by adding shapelets \citep{Refregier03,Birrer++15}, with the amount small-scale structure in the source controlled by the parameter $n_{\rm{max}}$, which we set to $n_{\rm{max}} = 8$. In the case of DESJ0408+5354, we follow \citet{Shajib++19b} and add additional set of shapelets with $n_{\rm{max}} = 3$ to one of the additional lensed sources. Adding complexity to the background source and fitting the lens model to the mock data removes the residuals in the image plane. This suggests the substructure-induced imaging residuals depend explicitly on the parameterization of the background light source. We elaborate on this topic in Appendix \ref{app:imagingresiduals}.
	
	\begin{figure}
		\includegraphics[clip,trim=0cm 0cm 0cm
		1cm,width=.48\textwidth,keepaspectratio]{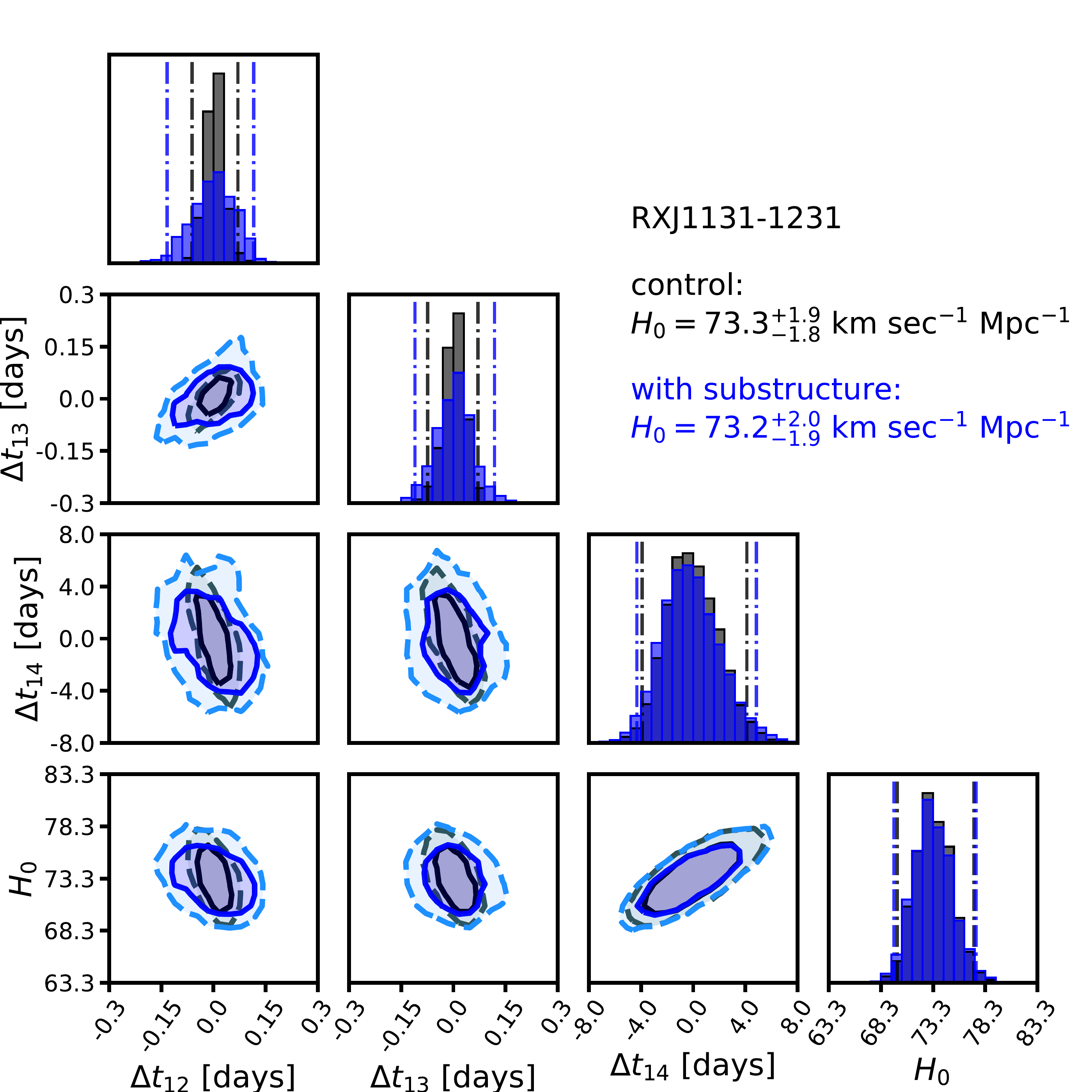}
		\caption{\label{fig:lens1131tdelayH0} The joint posterior distribution of the residual time delays $\Delta t$, defined as the model-predicted time delay minus the `true' time delay, and the Hubble constant, computed for the lens system RXJ1131-1231. The contours show the $68 \%$ and $95\%$ confidence intervals for the control posterior that includes no substructure (black), and the the combined posteriors from the inferences on lens models that include substructure (blue). Dashed lines in the histograms show 95$\%$ confidence intervals. The median and 68$\%$ confidence intervals of the $H_0$ inference are printed above the figure.}
	\end{figure}
	
	\begin{figure}
		\includegraphics[clip,trim=0cm 0.2cm 0cm
		1cm,width=.48\textwidth,keepaspectratio]{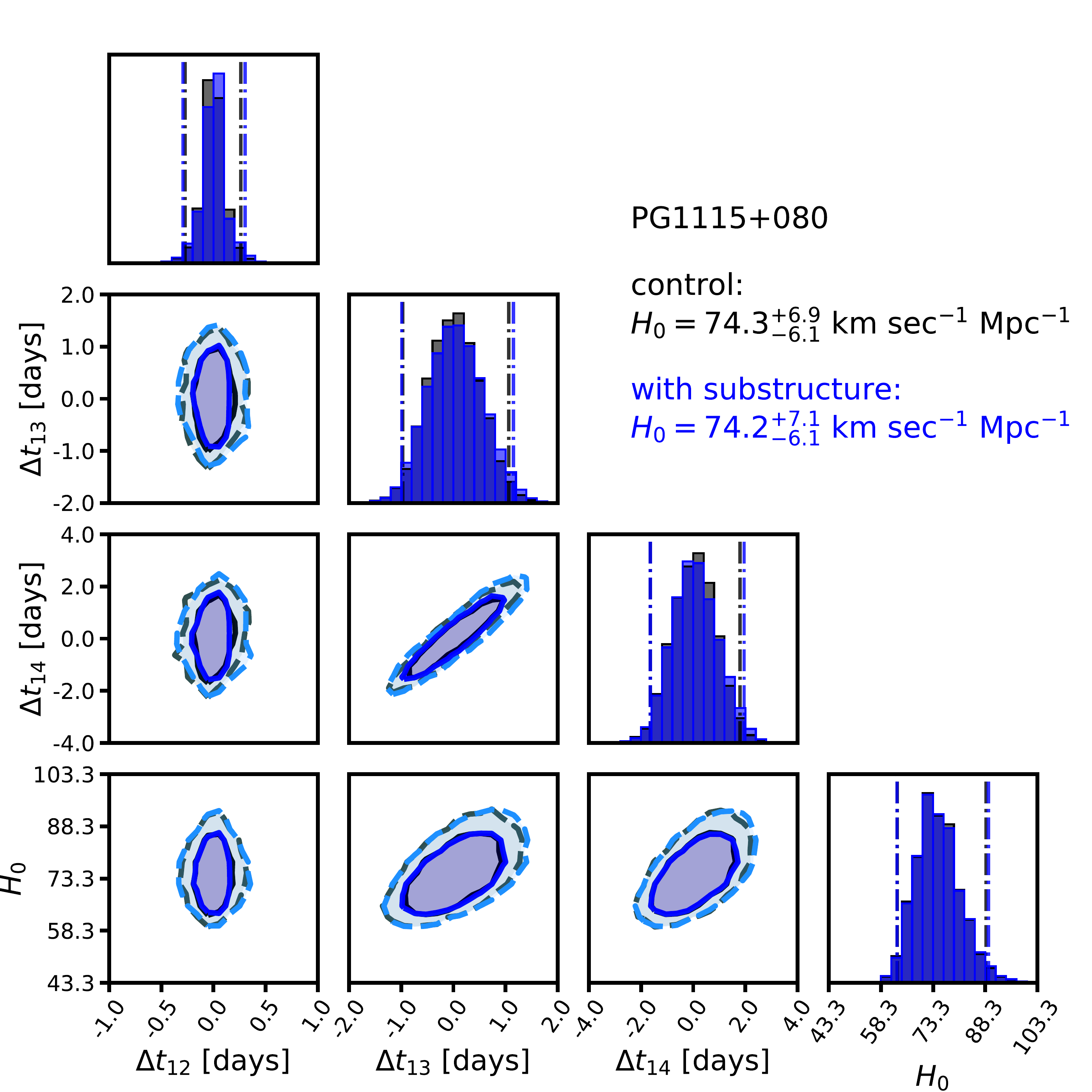}
		\caption{\label{fig:lens1115tdelayh0} The figure is the same as Figure \ref{fig:lens1131tdelayH0}, but shows the posteriors of the residual time delays and $H_0$ computed for the lens system PG1115+080.}
	\end{figure}
	
	\begin{figure}
		\includegraphics[clip,trim=0cm 0.2cm 0cm
		1cm,width=.48\textwidth,keepaspectratio]{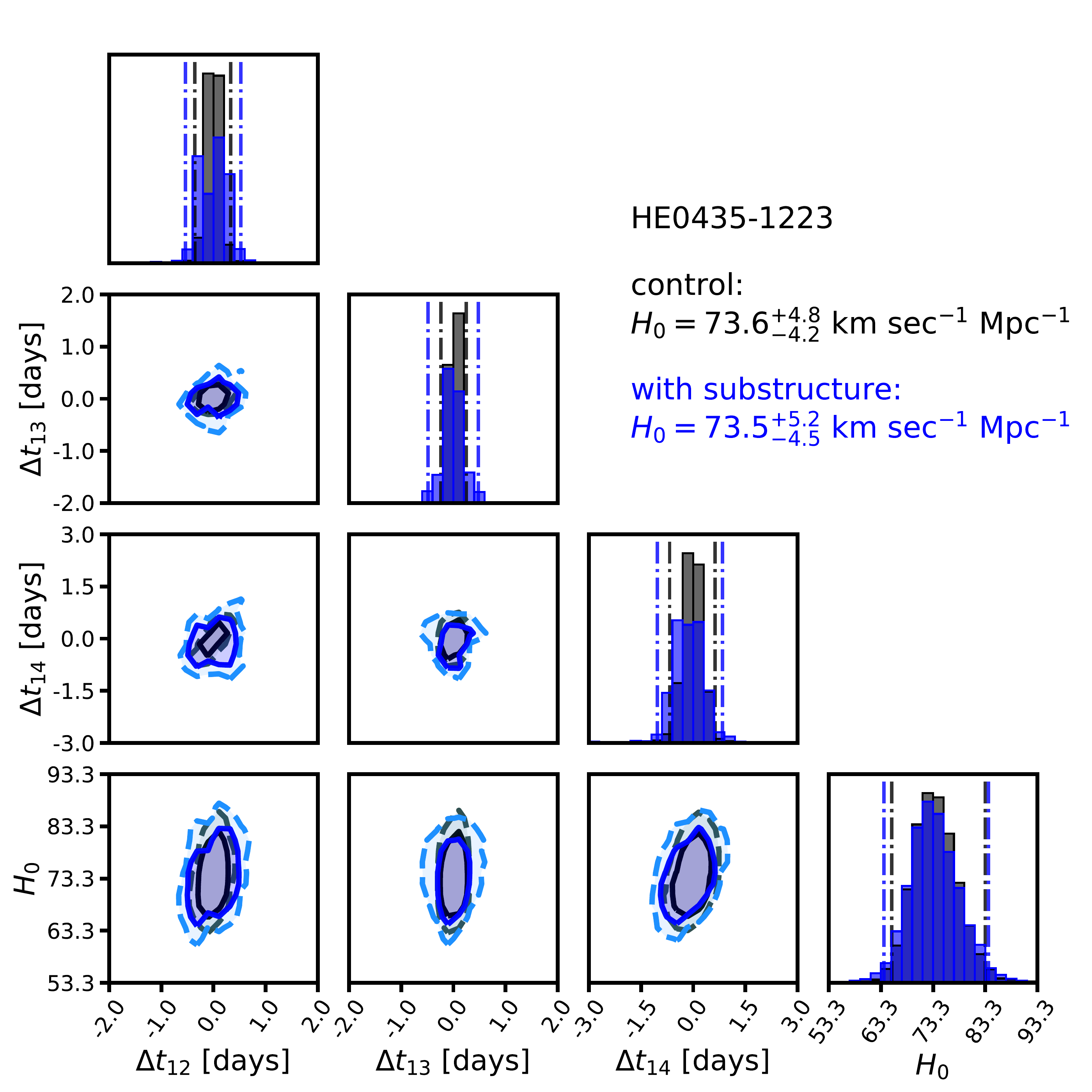}
		\caption{\label{fig:lens0435tdelayh0} The figure is the same as Figure \ref{fig:lens1131tdelayH0}, but shows the posteriors of the residual time delays and $H_0$ computed for the lens system HE0435-1223.}
	\end{figure}
	
	\begin{figure}
		\includegraphics[clip,trim=0cm 0.2cm 0cm
		1cm,width=.49\textwidth,keepaspectratio]{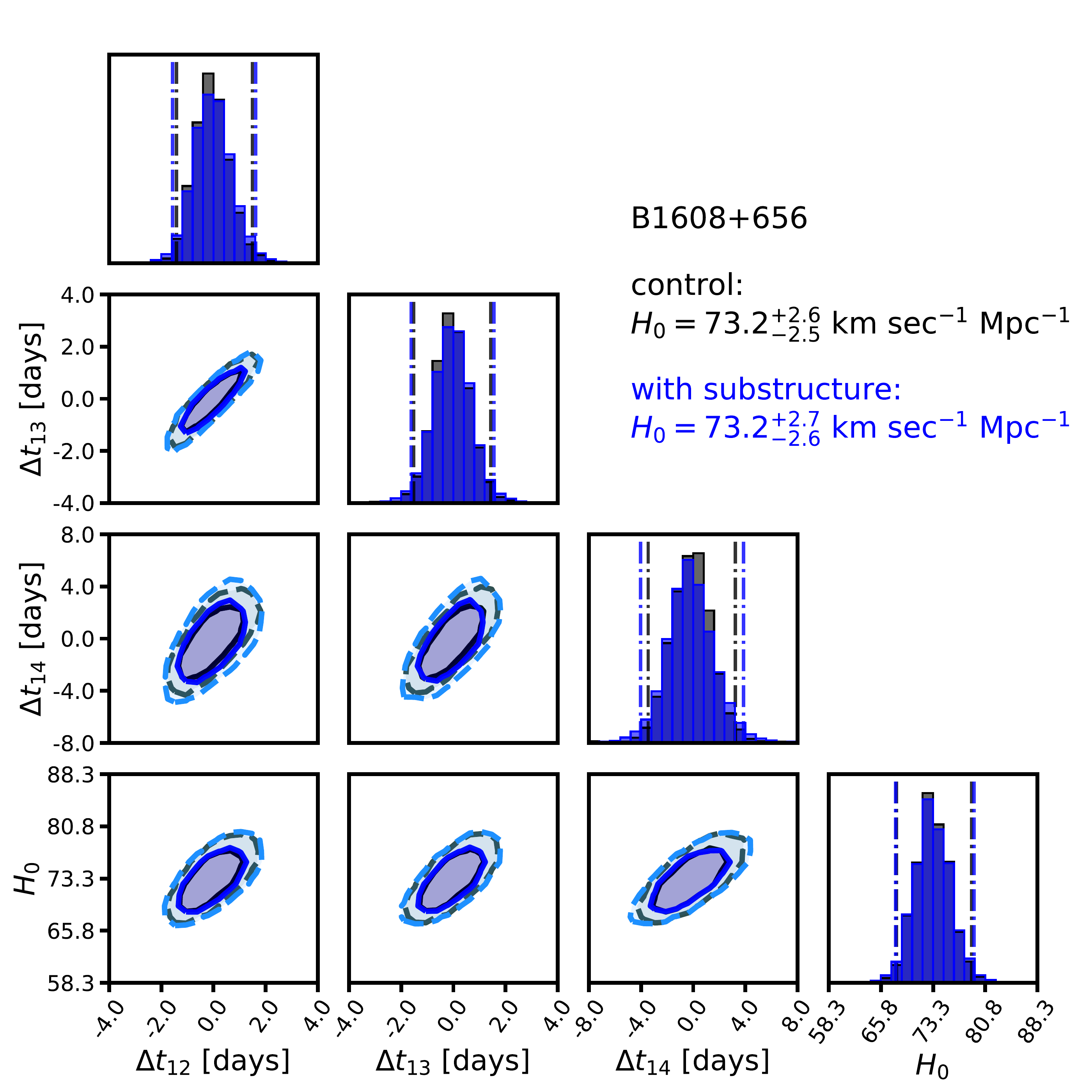}
		\caption{\label{fig:lens1608tdelayh0}  The figure is the same as \ref{fig:lens1131tdelayH0}, but shows the posteriors of the residual time delays and $H_0$ computed for the lens system B1608+656.}
	\end{figure}
	
	\begin{figure}
		\includegraphics[clip,trim=0cm 0.2cm 0cm
		0.5cm,width=.48\textwidth,keepaspectratio]{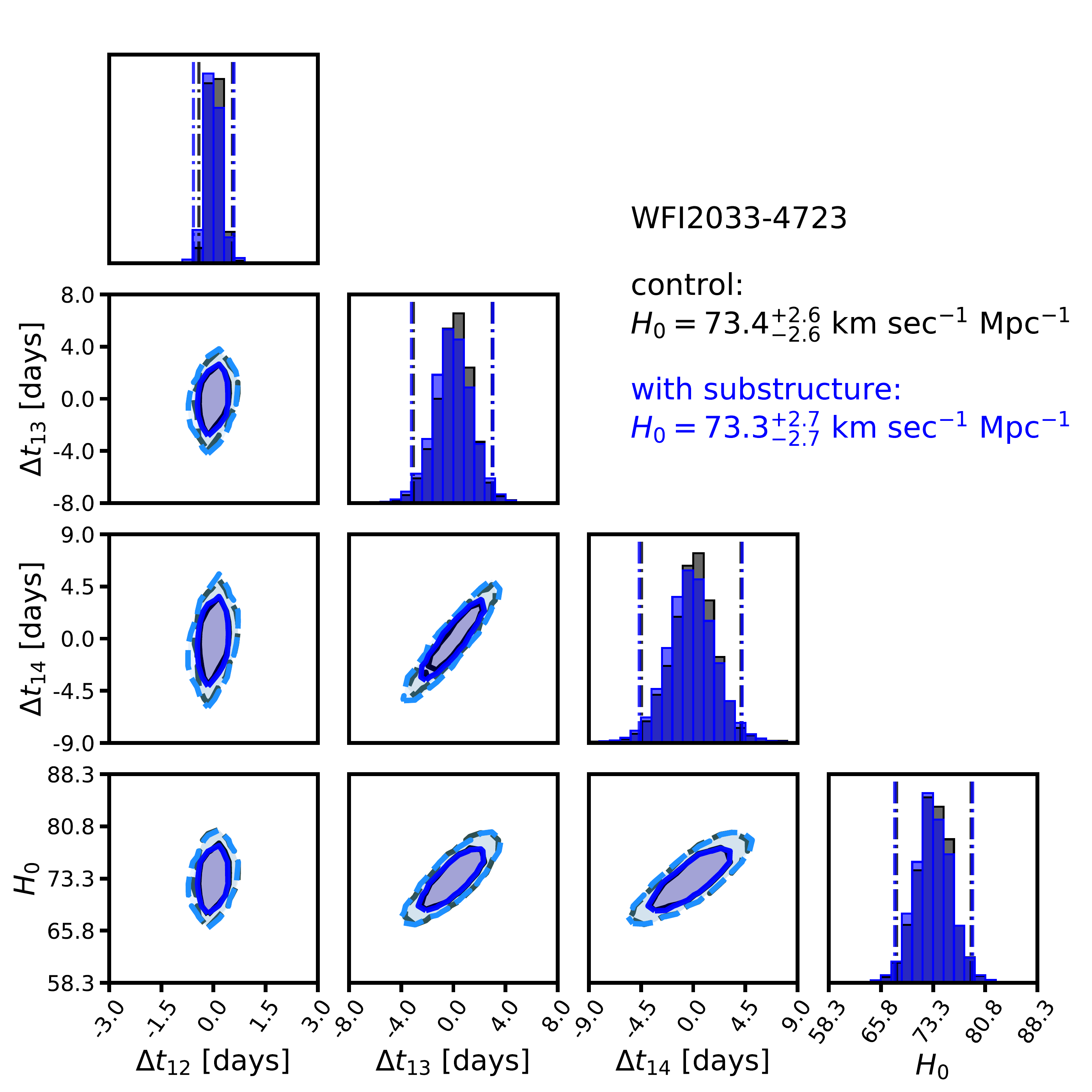}
		\caption{\label{fig:lens2033tdelayh0}  The figure is the same as \ref{fig:lens1131tdelayH0}, but shows the posteriors of the residual time delays and $H_0$ computed for the lens system WFI2033-4723.}
	\end{figure}
	
	\begin{figure}
		\includegraphics[clip,trim=0cm 0.2cm 0cm
		0.5cm,width=.48\textwidth,keepaspectratio]{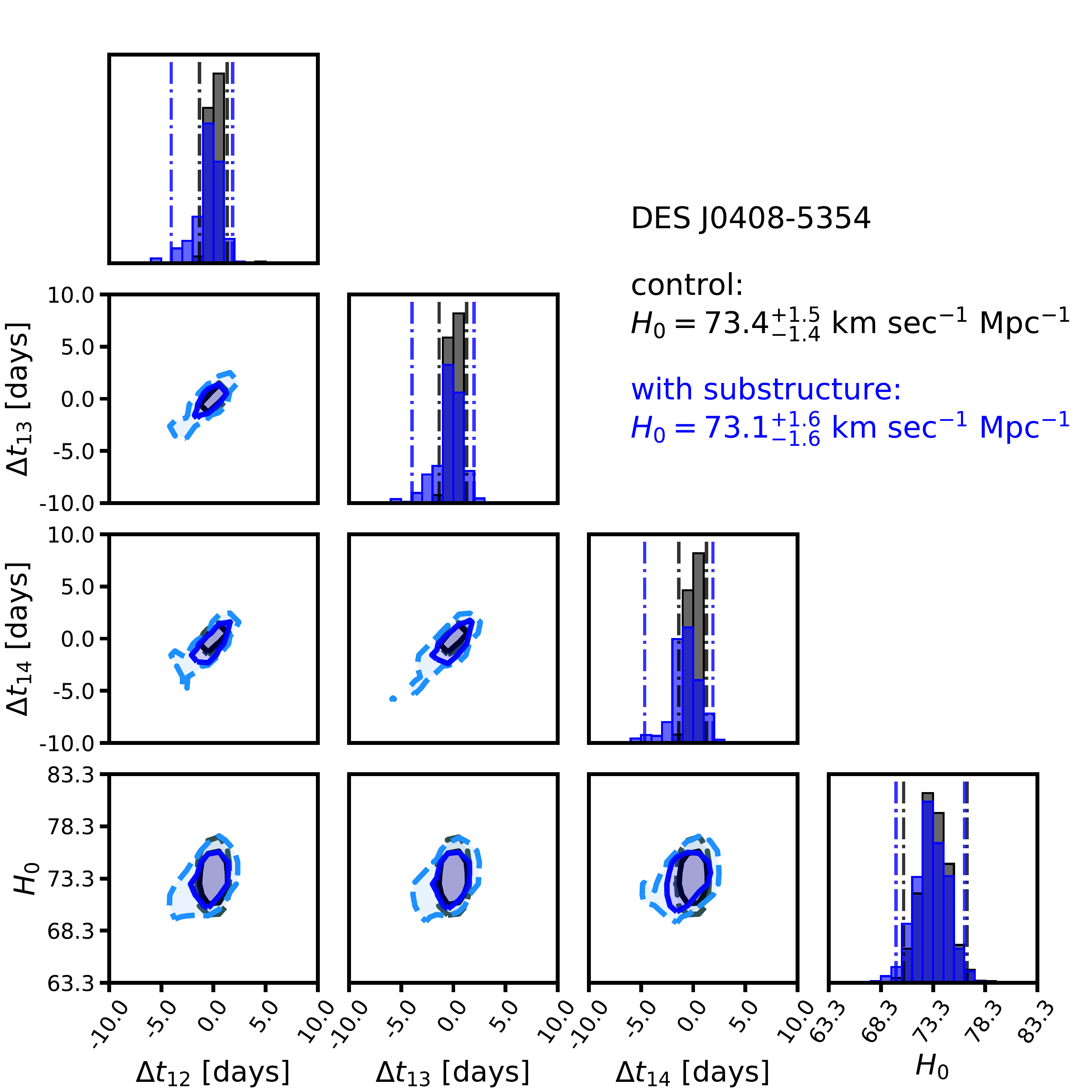}
		\caption{\label{fig:lens0408tdelayh0}  The figure is the same as \ref{fig:lens1131tdelayH0}, but shows the posteriors of the residual time delays and $H_0$ computed for the lens system DESJ0408-5354.}
	\end{figure}
	
	\subsection{Discussion of individual lens systems}
	\label{ssec:individualdis}
	We construct realistic analogs of six strong lens systems analyzed by the TDCOSMO collaboration. We briefly comment on each system in following paragraphs, highlighting both cosmographic inferences and substructure lensing analyses performed with each system.
	
	\begin{itemize}
		\item{RXJ1131-1231} This quad was discovered by \citet{Sluse++03}, and consists of source at $z = 0.658$ lensed by a galaxy at $z = 0.295$ \citep{Sluse++03,Sluse++07}. The time delays are measured at  $\sim$ 1.6$\%$ precision by \citet{Tewes++13}, and exhibit a relatively long maximum delay of $\sim 90$ days. This system has been modeled extensively with the aim of constraining cosmology \citep{Suyu++13,Suyu++14,Birrer++16,Chen++16,Chen++18}. The lens mass model includes a satellite galaxy inside the Einstein radius. Perturbations from substructure in the lensed arc of this system were analyzed by \citet{Birrer++17}, and were used to place a lower limit on the mass of a thermal relic dark matter particle of $2.2 \rm{keV}$. 
		
		\item{PG1115+080} This quad was discovered by \citet{Weymann++80}, and consists of a source at redshift $z = 1.722$ lensed by a galaxy at $z_d = 0.310$ \citep{Tonry++98}. The time delays were measured at $\sim 6.4\%$ precision by \citet{Courbin++18}, and the longest delay is $\sim 18$ days. This system was recently modeled with both adaptive optics and HST imaging by \citep{Chen++19}. We explicitly model the nearby cluster of galaxies as a single SIS profile projected $\sim 10 \ \rm{arcsec}$ away from the lens center. This system was first used to measure the abundance of dark subhalos by \citet{DalalKochanek02}, and more recently was used to constrain the free-streaming length of dark matter \citet{Hsueh++19,Gilman++20a}, and the mass-concentration relation of CDM halos \citep{Gilman++20b}. 
		
		\item{HE0435+1223} This quad was discovered by \citet{Wisotzki++02}, with lens/source redshifts of $z = 0.45$ and $z_s = 1.69$, respectively \citep{Morgan++05}.
		The time delays were measured at $\sim 5.3\%$ precision \citep{Bonvin++17}, with a maximum delay of $\sim 14$ days. This system was analyzed for cosmography by \citet{Bonvin++17,Wong++17,Chen++19}. We explicitly model the galaxy located behind the main deflector at $z = 0.78$ as an SIS profile, placing it at an angular position corrected by $\sim 0.65 \ \rm{arcsec}$ due to foreground lensing effects from the main deflector\footnote{For additional discussion, see Section 5.8 in \citet{Gilman++20a}}. \citet{Nierenberg++17} measured narrow-line flux ratios for this system \citep[see also][]{Nierenberg++19}, and used them to look for substructure near the images. The narrow-line flux ratio measurements for this system were also used measure the halo mass function and mass-concentration relation of CDM halos \citep{Gilman++20a,Gilman++20b}.
		
		\item{B1608+656} This lens system was discovered by \citet{Myers++96}, with time delay measurements and early cosmographic inferences carried out by \citet{Fassnacht++02} and \citet{Koopmans++03}. The lens/source redshifts for this system are $z_d = 0.45$, and $z_s = 1.69$ \citep{Fassnacht++96, Myers++96}. The longest delay in this system is $\sim 76$ days, measured at $\sim 2\%$ precision. This system is unique in that main deflector consists of two merging galaxies, although \citep{Suyu++09} show it is still well-described by a single power law profile. We follow \citet{Suyu++09} and model the larger object as a power law ellipsoid, and included smaller object as an SIS profile. 
		
		\item{WFI2033-4723} This quad was discovered by \citet{Morgan++04}. The lens and source lie at $z = 0.66$ and $z = 1.66$, respectively \citep{Sluse++19}. The time delays were measured by \citet{Bonvin++19} with a precision of $\sim 2\%$, and a maximum delay of $\sim 60$ days. The main lensing galaxy has a small satellite located $\sim 2 \ \rm{arcsec}$ away from the lens centroid. In addition, there is a galaxy located behind the main deflector at $z = 0.745$ situated $\sim 3.5$ arcseconds away from the lens center \footnote{As in the case of HE0435+1223, since this galaxy lies behind the main lens plane the coordinate listed in Table \ref{tab:baselinemodels} is corrected for foreground lensing effects.}. This system was modeled for cosmography by \citet{Rusu++19}, and for substructure lensing inferences by \citep{Gilman++20a,Gilman++20b}. 
		
		\item{DESJ0408-5354} This lens systems stands out from the rest due to the presence of two strongly-lensed sources, in addition to the quadruply-imaged quasar at $z = 2.355$, discovered by \cite{Lin++17}. The time delays for this system were measured by \citet{Courbin++18b} at a precision of $1\%$ with the longest delay $\sim 150$ days. 
		\citet{Shajib++19b} modeled this system for cosmography, and exploited the additional information provided by the multiple lensed sources to measure $H_0$ at 3.8$\%$ precision (not including substructure). The main deflector is accompanied by a satellite galaxy that splits one of the lensed images, and several other galaxies along the line of sight. \citet{Shajib++19b} modeled six of these line-of-sight galaxies, but in our mock lens models we only include the nearest two (G3 and G4, using the naming convention from \citet{Shajib++19b}). This simplification does not affect our conclusions, since we are matching the precision of the original inference by construction.  
		
		We do not explicitly model doubly-imaged quasar SDSS 1206+4332, which was analyzed for cosmography by \citet{Birrer++19a}, because the analysis pipeline we develop is designed for quadruple-image lenses. However, based on the results obtained for the six systems we do model, we can estimate the fractional uncertainty in $H_0$ from substructure in SDSS 1206+4332 (see Section \ref{ssec:h0}). 
		
	\end{itemize}
	
	The control lens models used for each system are summarized in Table \ref{tab:baselinemodels}. 
	
	\section{Results}
	\label{sec:results}
	
	In this section we apply the procedure outlined in Section \ref{sec:methodology} to the six strong lens systems described in the previous section. In Section \ref{ssec:maps}, we show examples of the mock and reconstructed lensed images for each system, as well as maps of the residual time delay surface and convergence. Section \ref{ssec:h0} presents the main results of this analysis, in which we quantify how un-modeled dark substructure contributes to uncertainties and/or bias in $H_0$. 
	
	\subsection{Reconstructed lenses: imaging residuals, time delay anomalies, and convergence}
	\label{ssec:maps}
	
	Figures \ref{fig:lensmaps1} and \ref{fig:lensmaps2} show examples of the mock observed and reconstructed lenses for each of the six systems we analyze. The panels on the far left show the mock observed lens system, which includes substructure. The reconstructed lens shown in the second from left panels includes no substructure. The central panels show the imaging residuals between the mock observed lens and the reconstruction, and displays the reduced $\chi^2$ obtained from the fit to the imaging data. We discard realizations with reduced $\chi^2 < -1.1$, as a structure massive enough to produced large imaging residuals would likely be detected and explicitly modeled. In Appendix \ref{app:imagingresiduals} we comment on the implications of these imaging residuals for gravitational imaging of dark substructure, and in particular on the implications for recent proposals to measure the convergence power spectrum of halos through gravitational imaging residuals.   
	
	The second from right panels show the residual time delay surface, which is the difference between the `true' time delay surface of the control dataset, and the model-predicted time delay surface from the reconstructed lens system. Dark matter halos imprint small perturbations on the time delay surface, which, if they were somehow detected, could be interpreted as `time delay anomalies', similar to the `flux ratio anomalies' attributed to dark substructure. 
	
	The far right panels of Figures \ref{fig:lensmaps1} and \ref{fig:lensmaps2} show the residual convergence\footnote{Convergence is typically defined as the projected mass in units of the critical density for lensing.} between the full lens system that includes substructure, and the smooth model fit to the mock data. We use the definition of an effective convergence in terms of the deflection angles $\boldsymbol{\alpha}$ \citep{Gilman++19a} 
	\begin{equation}
	\label{eqn:multiplaneconv}
	\kappa_{\rm{eff}} = \frac{1}{2} \left(\frac{\partial \boldsymbol{\alpha}}{\partial \boldsymbol{x}} + \frac{\partial \boldsymbol{\alpha}}{\partial \boldsymbol{y}}\right),
	\end{equation}
	which generalizes lensing convergence to systems with line-of-sight structure\footnote{The deflection angle $\boldsymbol{\alpha}$ is just the summation term to the right of the minus sign in Equation 
		\ref{eqn:raytracing}}. The residual convergence shown in the right-most panels of Figures \ref{fig:lensmaps1} and \ref{fig:lensmaps2} is simply $\kappa_{\rm{eff}} - \kappa_{\rm{reconstructed}}$, where $\kappa_{\rm{reconstructed}}$ is the convergence from the smooth model fit to the mock data and $\kappa_{\rm{eff}}$ is computed from the mock lens system that includes substructure. Structures in the convergence maps that are tangentially sheared around the Einstein radius come from halos located between the main deflector and the source that produce highly nonlinear lensing effects. 
	
	\subsection{$H_0$ inferences with dark substructure}
	\label{ssec:h0}
	Figures \ref{fig:lens1131tdelayH0}, \ref{fig:lens1115tdelayh0}, \ref{fig:lens0435tdelayh0}, \ref{fig:lens1608tdelayh0}, \ref{fig:lens2033tdelayh0} and \ref{fig:lens0408tdelayh0} show the joint posterior distribution for the inferred value of $H_0$ and the residual time delays $\Delta t$, defined as the true time delay from the mock data minus the model-predicted time delay. The posterior from the control data is shown in black, and the posterior obtained from the substructure-perturbed datasets are shown in blue. The joint posterior distributions of $H_0$ and the lens macromodel parameters are shown in Appendix \ref{sec:apph0macro}. 
	
	\begin{table*}
		\centering
		\caption{Summary of the sources of uncertainty affecting cosmographic inferences. This table is the same as Table 3 in the paper by \citet{Wong++19}, plus the additional uncertainty term from substructure.}
		\label{tab:results}
		\renewcommand{\arraystretch}{1.2}
		\setlength{\tabcolsep}{5pt}
		\begin{tabular}{lcccccccr} 
			\hline
			source of uncertainty & RXJ1131-1231 & PG1115+080 & HE0435-1223 & B1608+656 & WFI2033-4723 & DESJ0408-5354 \\ 
			\hline	
			time delays & 1.6\%& 6.4\%& 5.3\%& 1.7\%& 2.9\%& 1.8\% \vspace{0.05in}\\ 
			external convergence & 3.3\%& 2.7\%& 2.8\%& 6.4\%& 5.7\%& 3.3\%\vspace{0.05in}\\ 
			lens modeling & 2.2\%& 5.7\%& 2.5\%& 3.0\%& 2.2\%& 1.0\%\vspace{0.05in}\\  
			substructure & 0.7\%& 1.8\%& 2.4\% & 0.9\%& 1.2\%& 1.1\% \vspace{0.05in}\\  
			\hline
			$\sigma_{\rm{reported}}$ & 4.3\%& 9.0\%& 6.5\%& 7.3\%& 6.8\%&3.9\% \\
			$\sigma_{\rm{total}}$ & 4.4\% & 9.2\%& 6.9\%& 7.4\%& 6.9\%&  4.1\%\vspace{0.05in}\\ 
			\hline 
		\end{tabular}
	\end{table*}
	
\begin{figure}
		\includegraphics[clip,trim=0.5cm 0.4cm 0.4cm
		0.5cm,width=.45\textwidth,keepaspectratio]{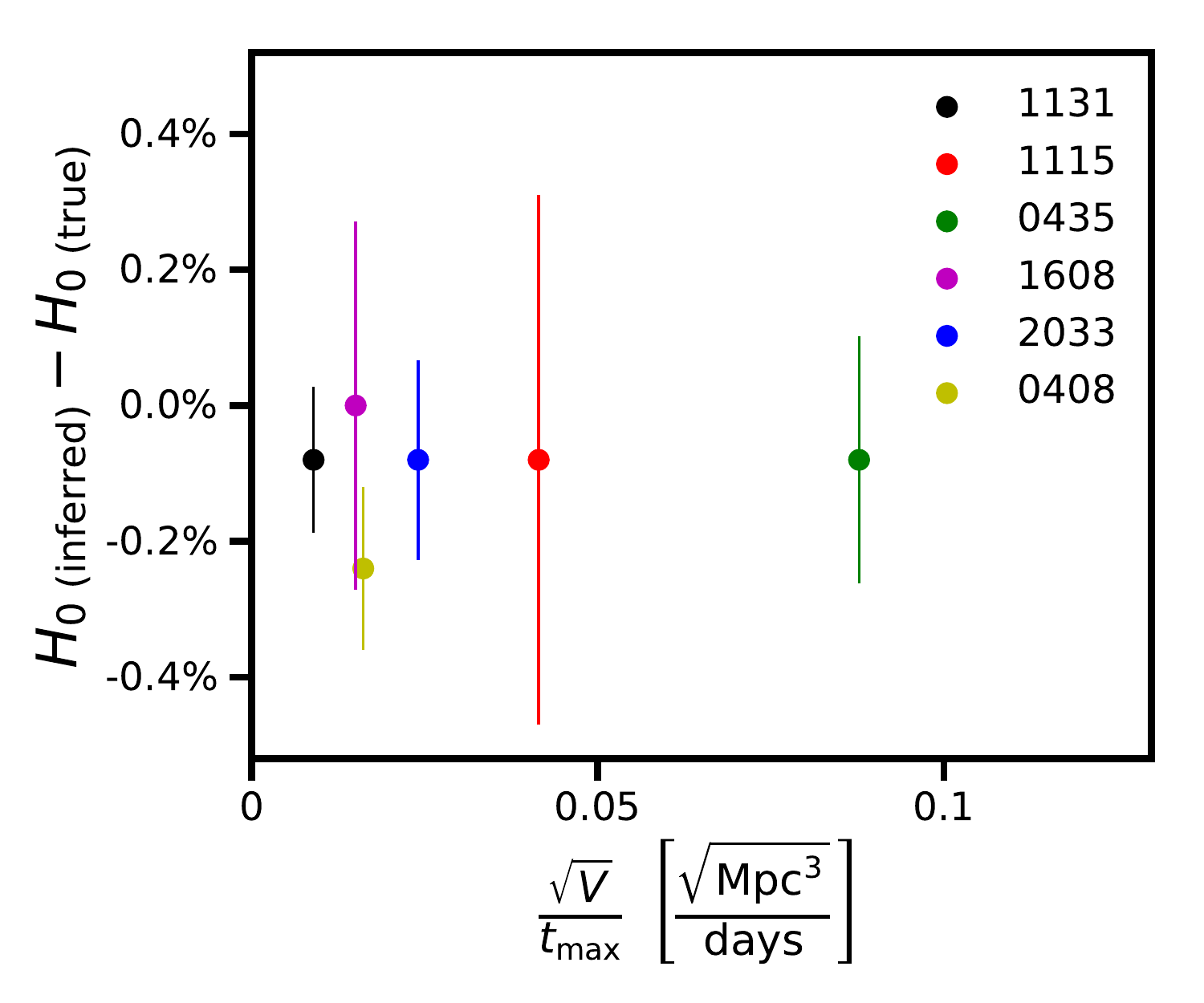}
		\caption{\label{fig:nobias} The y-axis shows the bias in $H_0$, defined as the median of the $H_0$ posterior obtained from the substructure-perturbed datasets minus the median of the posterior from the control datasets. The mock lenses are sorted on the x-axis by the square root of the lensing volume divided by the longest time delay, a dimensional quantity that determines the additional uncertainty caused by substructure in the lens system. The error bars show the statistical uncertainty on the median, which is given by $\frac{\sigma_{\rm{control}}}{\sqrt{N}}$, where $\sigma_{\rm{control}}$ is the uncertainty in $H_0$ from the control dataset and $N=200$ is the number of realizations. We find no evidence for bias in the inferred value of $H_0$ stemming from substructure that is not modeled in cosmographic inferences.}
	\end{figure}

	The median inferred value of $H_0$ can differ from the `true' value of $H_0 = 73.3 \ \rm{km} \ \rm{s}^{-1} \ \rm{Mpc^{-1}}$ if the $H_0$ posterior is asymmetric, as is the case for PG1115+080. This does not suggest the presence of a bias in the inference, because a bias would manifest as different medians between the control and substructure-perturbed posteriors. We plot the difference between the medians of the two posteriors as a function of the square root of the lensing volume divided by the longest time delay (the significance of these units will be discussed in the following paragraph) in Figure \ref{fig:nobias}. The median values of the two posteriors are statistically consistent to within the error bars, which are given the uncertainty of the control dataset divided by $\sqrt{N}$, where $N=200$ is the total number of realizations. We conclude that the omission of substructure from strong lens models used for cosmography does not introduce a bias greater than 0.3$\%$. 
	
	\begin{figure}
		\includegraphics[clip,trim=0.25cm 0.25cm 0.25cm
		0.25cm,width=.45\textwidth,keepaspectratio]{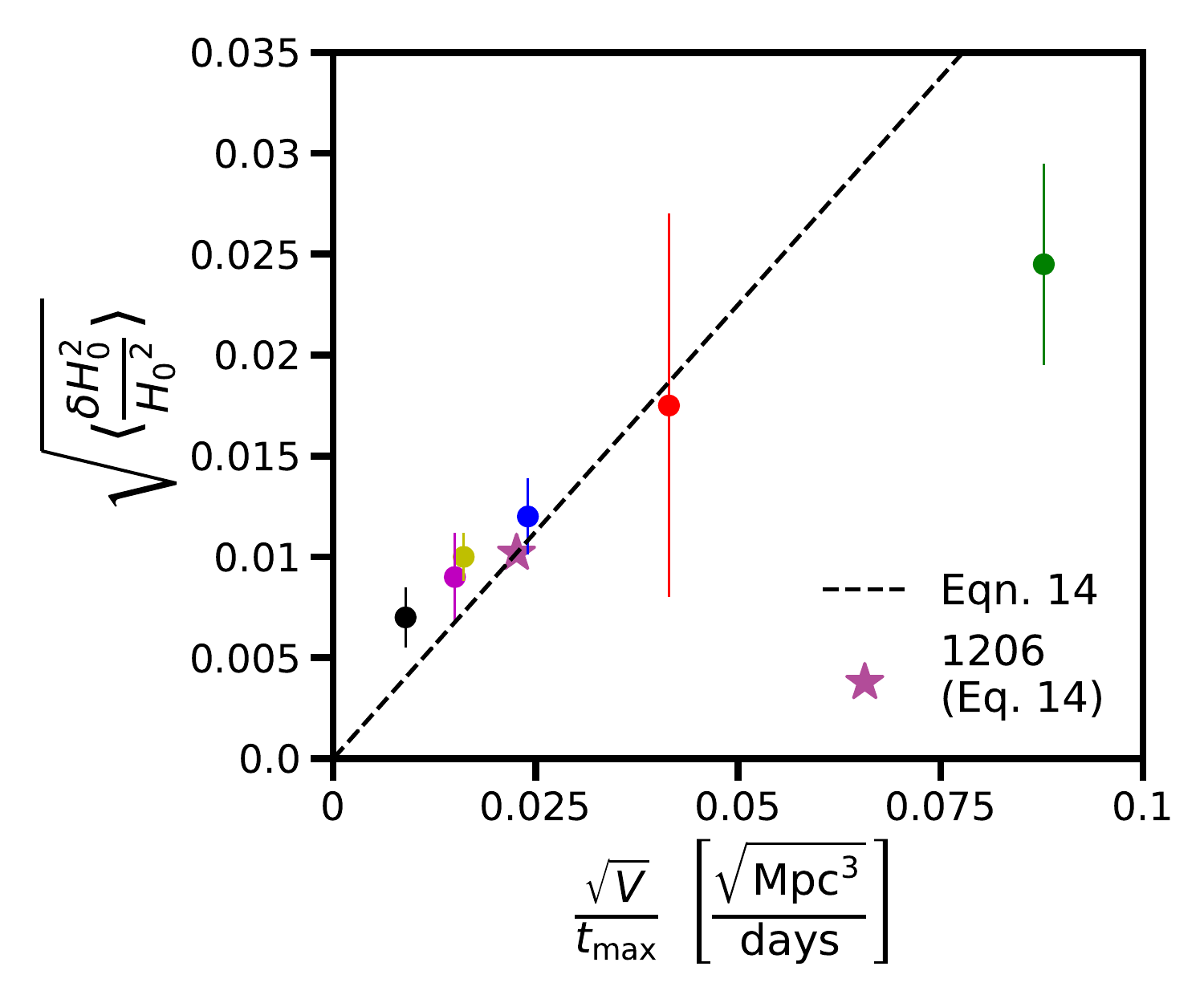}
		\caption{\label{fig:redshiftevo} The r.m.s. fractional uncertainty in $H_0$ (y-axis) is plotted as a function of the number of halos divided by the longest time delay, which is proportional to the lensing volume divided by the longest time delay (x-axis). The color labeling of the points is the same as in Figure \ref{fig:nobias}. The points show the fractional uncertainty for each of the six systems analyzed in this work, and error bars show the $68 \%$ bootstrap confidence intervals. The dashed line shows the fitting function given by Equation \ref{eqn:fittingfunc}. We use the fitting function in Equation \ref{eqn:fittingfunc} to estimate the contribution of substructure to the uncertainty in $H_0$ for the lens system SDSS 1206+4332 (purple star).}
	\end{figure}
	
	While we find no evidence for a bias in $H_0$, substructure alters the lens model and changes the Fermat potential at the image positions, which leads to an increased scatter in the time delays that factors directly into the $H_0$ error budget. The additional uncertainty caused by substructure evolves with redshift and geometry in a predictable way. From Equation \ref{eqn:tabmulti}, the fractional uncertainty in $H_0$ scales as
	\begin{equation}
	\frac{\delta H_0}{H_0} \propto \frac{\delta t}{t_{\rm{max}}}
	\end{equation}
	where $t_{\rm{max}}$ is the longest time delay, which is usually the most precisely measured and therefore the most constraining observable, and $\delta t$ is a perturbation to the time delays from dark matter halos. Using analytic arguments, \citet{KeetonMoustakas09} and \citet{Cyr-Racine++16} showed that $\delta t$ is proportional to $\sqrt{f_{\rm{sub}}}$, where $f_{\rm{sub}}$ is proportional to the total number of subhalos. We can generalize these results to realizations with line-of-sight halos by replacing $f_{\rm{sub}}$ by $N_{\rm{halo}}$, where $N_{\rm{halo}}$ is just the total number of halos. The total number of halos is proportional to the lensing volume $V$, so the scaling in the fractional uncertainty in $H_0$ is
	\begin{equation}
	\label{eqn:trend}
	\frac{\delta H_0}{H_0} \propto \frac{\sqrt{V}}{t_{\rm{max}}}.
	\end{equation}
	The volume $V$ is defined in terms of the comoving distances to the lens $d_{\rm{lens}}$, to the source $d_{\rm{src}}$, and the Einstein radius (converted to radians) $R_{\rm{Ein}}$
	\begin{equation}
	V = \pi \left(d_{\rm{lens}} R_{\rm{Ein}}\right)^2 d_{\rm{src}}.
	\end{equation} 
	
	Figure \ref{fig:redshiftevo} shows the trend predicted by Equation \ref{eqn:trend} alongside the fractional uncertainties obtained for each lens. The dashed line fit in Figure \ref{fig:redshiftevo} is given by
	\begin{equation}
	\label{eqn:fittingfunc}
	\begin{split}
	\frac{\delta H_0}{H_0} = 0.045 \left( \frac{V}{1 \  \rm{Mpc^3}}\right)^{0.5} \left(\frac{10 \  \rm{days}}{t_{\rm{max}}}\right). 
	\end{split}
	\end{equation}
	This simple relation captures the leading-order effects of substructure, but there are additional factors that could affect the uncertainties and produce scatter around the $\frac{\sqrt{V}}{{t_{\rm{max}}}}$ curve. For example, \citet{Oguri++07} noted that lenses with symmetric image configurations are more sensitive to gravitational potential perturbations than lenses with asymmetric image configurations. Also, the Sheth-Tormen halo mass function in the range $10^6 - 10^9 M_{\odot}$ has a mild redshift dependence that we do not account for in the fitting function.
	
	We can use Equation \ref{eqn:fittingfunc} to estimate the fractional uncertainty in $H_0$ given the lens redshift, the source redshift, the Einstein radius, and the longest time delay. For the doubly-imaged system SDSS 1206+4332, Equation \ref{eqn:fittingfunc} gives an uncertainty of $1.0 \%$. This system was not analyzed explicitly in this work since the analysis pipeline is designed for quadruple-image lenses, but the physical arguments that motivate the scaling $\frac{\sqrt{V}}{\rm{t_{\rm{max}}}}$ should still apply to SDSS 1206+4332. 
	
	We summarize the total error budget for each lens system, including the additional source of uncertainty from dark substructure, in Table \ref{tab:results}. For a joint inference from a sample of lenses, we compute the contribution to the error budget from substructure $\sigma_{\rm{net}}$, given by $\sigma_{\rm{net}}^2 = \left(\sum {\sigma_i^{-2}}\right)^{-1}$, where $\sigma_i$ are the uncertainties from substructure of each lens system included in the joint inference. Based on the results for the lenses we explicitly model summarized in Table \ref{tab:results}, plus the $1.0\%$ uncertainty estimated using Equation \ref{eqn:fittingfunc} for SDSS 1206+4332, we find $\sigma_{\rm{net}} = 0.5 \%$. This is an independent source of uncertainty that should be summed in quadrature with other contributions. 
	
	\section{Summary}
	\label{sec:summary}
	We have carried out an analysis of how dark matter subhalos and field halos along the line of sight affect inferences of the Hubble constant $H_0$ obtained through time delay cosmography. Using mock lenses based on six quadruply-imaged quasars modeled by the TDCOSMO collaboration, we quantify on a case-by-case basis how dark substructure would have affected the precision and accuracy of cosmographic inferences from each system, had substructure been included in the model. Our main results are summarized as follows:
	
	\begin{itemize}
		\item We show that excluding dark substructure from lens models used to infer cosmological parameters does not bias inferences on $H_0$ at the level of the $0.3\%$, the numerical and statistical precision of our work. This rules out the possibility that explicitly modeling substructure in cosmographic inferences would alleviate the $H_0$ tension between early and late-Universe probes. It also suggests that substructure cannot explain the anticorrelation between the lens redshift and the inferred time delay distance noted by \citet{Wong++19}. \\
		\item Substructure perturbs the image arrival time delays and the global mass model, which translates into uncertainties on the value of $H_0$ derived from the time delays and the Fermat potential predicted by the lens model. The additional uncertainties from substructure listed in Table \ref{tab:results} range from $0.7\%$ in the case of RXJ1131-1231 to $2.4\%$ in the case of HE0435-1223. This source of uncertainty is an independent contribution to the error budget that should be added in quadrature with other contributions. This addition source of uncertainty does not significantly alter the current error budget per lens, but it will not be completely negligible as the precision per system improves, e.g. with the addition of spatially resolved kinematics \citep{Shajib++18}. Combining our results for the uncertainties in single lenses, the overall contribution of substructure to the error budget of the TDCOSMO sample, which should be added in quadrature with the other sources of uncertainty, is $0.5\%$. \\
		\item We show that the perturbation to $H_0$ scales as the square root of the lensing volume divided by the longest time delay. We provide a simple fitting function that uses this scaling to predict the level of perturbation for a strong lens system given the lens and source redshifts, the Einstein radius, and the longest delay.  
	\end{itemize}
	
	\section{Discussion and conclusions}
	\label{sec:conclusions}
	
	Our results are consistent with previous analysis that show $H_0$ inferences are internally consistent \citep{Millon++19,Wong++19}, and are consistent with cosmological distances inferred from Type Ia supernova \citep{Pandey++19}. Since the level of perturbation from line-of-sight halos evolves with redshift, if un-modeled halos biased $H_0$ inferences significantly it would lead to inconsistencies between lenses at different redshifts, and between strong-lensing based inferences on $H_0$ with independent probes of the distance ladder. 
	
	While we find no evidence for bias, we conclude that the omission of substructure from strong lens models results in a slight underestimate of the uncertainties in the measured values of $H_0$. The additional uncertainty term, which should be summed in quadrature with the other contributions to the error budget, ranges from less than one percent in the case of RXJ1131-1231, to $2.4 \%$ in the case of HE0435-1223, increasing the total uncertainties in these systems from $4.3\%$ to $4.4\%$ and $6.5\%$ to $6.9\%$, respectively. When inferences from the seven lenses analyzed by TDCOSMO are combined, we estimate that substructure contributes an additional 0.5$\%$ uncertainty. This is a subdominant contribution to the overall error budget that we reiterate should be summed in quadrature with other contributions. 
	
	To make use of the results presented in this paper for future cosmographic inferences, we suggest two possibilities. First, one could obtain a rough estimate of the level of perturbation using the fitting function plotted in Figure \ref{fig:redshiftevo}, and possibly incorporate this additional uncertainty when quoting results on $H_0$. Second, the analysis performed in this work could be repeated for new lens systems. The first approach sacrifices precision for speed, while the second approach is more time consuming, but gives a more precise estimate for the additional uncertainty from substructure. 
	
	Finally, we note that dark matter halos should also alter the arrival times between images of lensed supernova, such as supernova ``Refsdal", which is multiply imaged by an early-type galaxy in the cluster MACS J1149.5+2223 \citep{Kelly++15}. We can extrapolate the results of this work to the Refsdal system to approximate the impact of substructure on future cosmographic inferences. The Refsdal system consists of four images arranged in an Einstein cross configuration separated by $\sim 1^{\prime \prime}$ with a maximum delay of $24$ days \citep{Rodney++16}, and a fifth image $\sim 30^{\prime \prime}$ away with a time delay of $\sim 1$yr. Coincidentally, both pairs of Einstein radii and time delays give an uncertainty of $2.1 \%$ in $H_0$ from substructure, per Equation \ref{eqn:fittingfunc}, because the long time delay between $S_X$ and the Einstein cross counteracts the larger image separation.  
	
	\section*{Acknowledgments}
	We thank Anowar Shajib for answering many questions, and for sharing the MCMC chains computed for DESJ0408-5354. We also thank Adriano Agnello, Frederic Courbin, and Dominique Sluse for comments on the manuscript, and the anonymous referee for constructive feedback. 
	
	DG, TT, and SB acknowledge support by the US National Science Foundation through grant AST-1714953 and AST-1906976. This work used computational and storage services associated with the Hoffman2 Shared Cluster provided by the UCLA Institute for Digital Research and Education's Research Technology Group. 
	\bibliographystyle{aa}
	\bibliography{bibliography}
	
	\appendix
	
	\section{Baseline macro lensmodels}
	\label{sec:appA}
	Table \ref{tab:baselinemodels} lists the  parameters describing the lens macromodels for the mock lens analogs of the six strong lens systems considered in this work.

	\begin{table*}
		\centering
		\caption{Lens model parameters used to create the baseline datasets. The main deflector is modeled as an elliptical power law plus external shear, and other satellites or field galaxies are modeled with spherical power law profiles. Parameters from left to right: Einstein radius of the main deflector, x and y mass centroid of main deflector, main deflector ellipticity and position angle, logarithmic slope of the main mass profile, external shear magnitude and position angle. The columns labeled $R_i$, $x_i$, $y_i$, $z_i$ contain the Einstein radii, coordinates, and redshifts of satellite galaxies, or galaxies along the line of sight, that are included in the mock lens models. Angular positions marked with an asterisk indicate that this coordinate  is corrected for foreground lensing effects by the main deflector.}
		\label{tab:baselinemodels}
		\renewcommand{\arraystretch}{1.2}
		\setlength{\tabcolsep}{3.pt}
		\begin{tabular}{lccccccccccccccccr} 
			\hline
			Lens system &$R_{\rm{Ein}}$ & $g_x$ & $g_y$ & $\epsilon$ & $\theta_{\epsilon}$ & $\gamma$ & $\gamma_{\rm{ext}}$ & $\theta_{\rm{ext}}$ & $R_1$ & $x_1$ 
			& $y_1$ & $z_1$ & $R_2$ & $x_2$ & $y_2$& $z_2$ \\
			\hline	
			RXJ1131-1231 & 1.58 & -0.01 & -0.03 & 0.14 & 34 & 1.98 & 0.12 & -82 & 0.28 & -0.1 & 0.61 & $z_d$ & - & - & - & - \\
			PG1115-080 & 1.05 & 0.0 & 0.0 & 0.01 & -71 & 2.20 & 0.03 & -56 & 2.0 & -9.21 & -3.91 & $z_d$ & - & - & - & - \\
			HE0435-1223 & 1.17 & -0.02 & 0.02 & 0.07 & -81 & 1.98 & 0.05 & 10 & 0.35 & $-2.27^{*}$ & $1.98^{*}$ & $z_d + 0.33$ & - & - & - & - \\
			B1608+656 & 0.90 & 0.01 & 0.06 & 0.51 & -4 & 2.08 & 0.10 & 26 & 0.26 & -0.71 & 0.13 & $z_d$ & - & - & - & - \\
			WFI2033-4723 & 1.0 & -0.02 & 0.02 & 0.22 & 57 & 1.95 & 0.17 & -8 & 0.03 & 0.24 & 2.04 & $z_d$ & 0.93 & $-3.63^{*}$ & $-0.68^{*}$ & $z_d + 0.09$ \\
			DESJ0408-5354 & 1.72 & 0.14 & 0.00 & 0.39 & 23 & 1.92 & 0.11 & 5 & 0.22 & -1.58 & -0.95 & $z_d$ & 0.77 & $1.13^{*}$ & $-7.45^{*}$ & $z_d + 0.17$ \\		
		\end{tabular}
	\end{table*}
	
	\section{Hiding dark matter halos in extended images with shapelets}
	\label{app:imagingresiduals}
	As a byproduct of this analysis, we obtain hundreds of mock lenses with realistic line-of-sight and substructure populations that we model with state of the art image reconstruction techniques. These simulations are relevant to the field of gravitational imaging of dark substructure in lensed arcs \citep{Vegetti++10,Hezaveh++16b,Vegetti++18}. Dark substructure occasionally produces imaging residuals in lensed arcs that betray the presence of individual halos, or provide a means to infer quantities such as the convergence power spectrum of halos in the lens system \citep{Hezaveh++16,DiazRivero++18,CyrRacine++19,Sengul++20}.
	
	As the convergence itself is not observable, it must be derived from the surface brightness residuals from the lens modeling \citep{CyrRacine++19}. Keeping all else fixed, adding additional complexity to the background source light partially removes gravitational imaging residuals in extended images. The gravitational imaging residuals therefore depend explicitly on how the unknown background source is modeled.  
	
\begin{figure*}
		\includegraphics[clip,trim=0.5cm 2cm 0.5cm
		3cm,width=.95\textwidth,keepaspectratio]{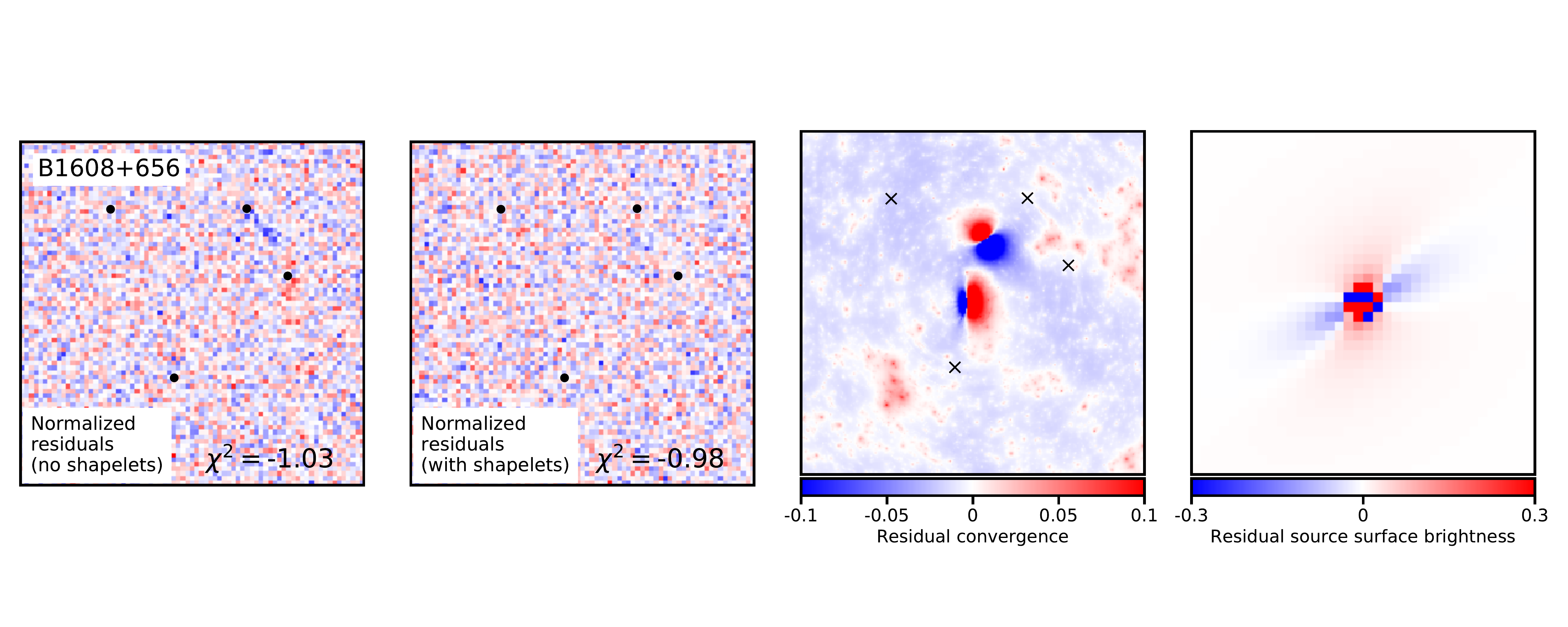}
		\includegraphics[clip,trim=0.5cm 2cm 0.5cm
		3cm,width=.95\textwidth,keepaspectratio]{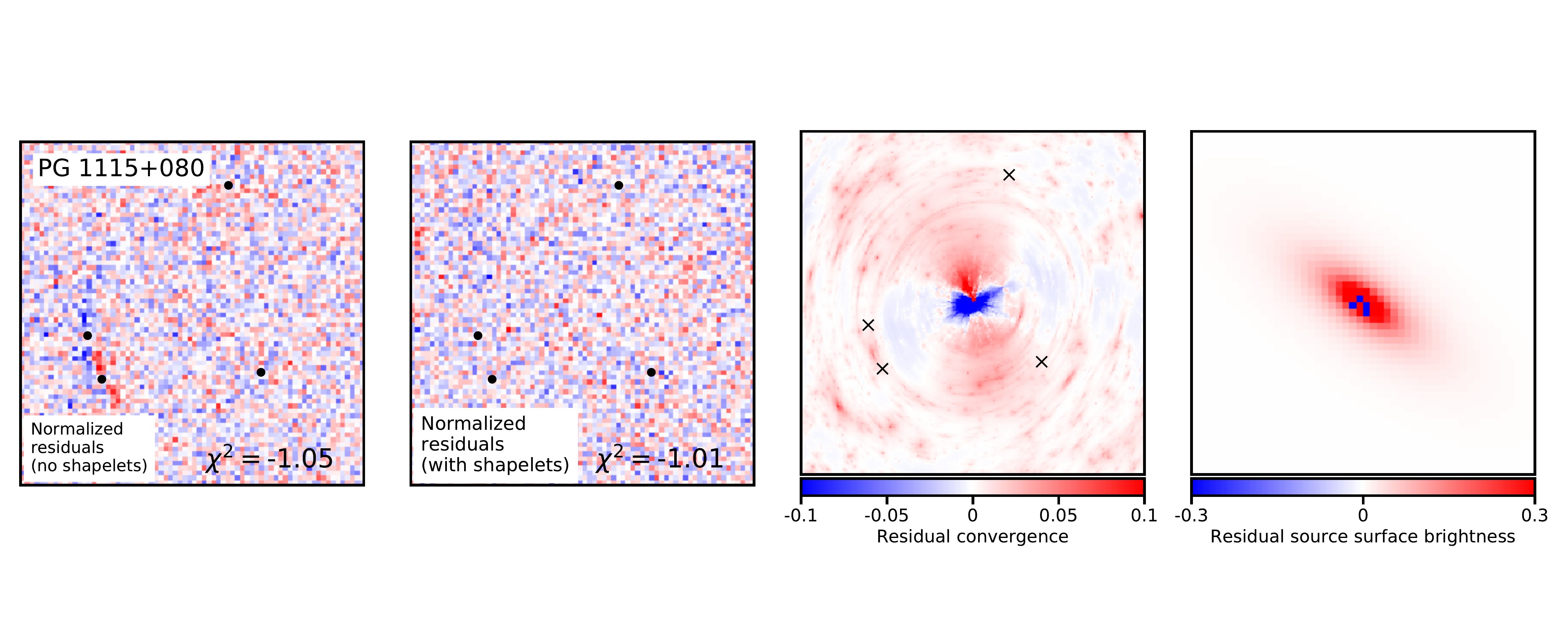}
		\caption{\label{fig:b1608shapvsnoshap}  The panels show examples of imaging residuals, convergence, and residual source surface brightness for one mock lens system based on B1608+656 (top row) and PG1115+080 (bottom row). The far left panel shows the gravitational imaging residuals when only an elliptical S\'{e}rsic profile, the same profile used to create the mock data, is used to model the background source. The second from left panel shows the imaging residuals when shapelets are added to the source light profile. The dark matter halos that perturb the lensed arc in the mock data are shown in the third from left panel, and the residual source surface brightness is shown in the far right panel. }
	\end{figure*}

	To illustrate, the two rows in Figure \ref{fig:b1608shapvsnoshap} show examples of imaging residuals computed for the mock analogs of B1608+656 and PG1115+080. The far left panels in Figure \ref{fig:b1608shapvsnoshap} show the imaging residuals without shapelets added to the background source, and the second from left panels show the same mock lens system, with the same population of dark matter halos, modeled with shapelets included in the background source. The population of line-of-sight halos and subhalos that perturbs the mock images are shown in the third panels, using the definition of the multi-plane convergence described in Section \ref{ssec:maps}. The far right panel shows the difference between the source surface brightness between the models that include and exclude shapelets. In these two cases, adding shapelets to the background source removes imaging residuals from the extended images. 
	
\begin{figure*}
		\includegraphics[clip,trim=0cm 0cm 0cm
		0cm,width=.45\textwidth,keepaspectratio]{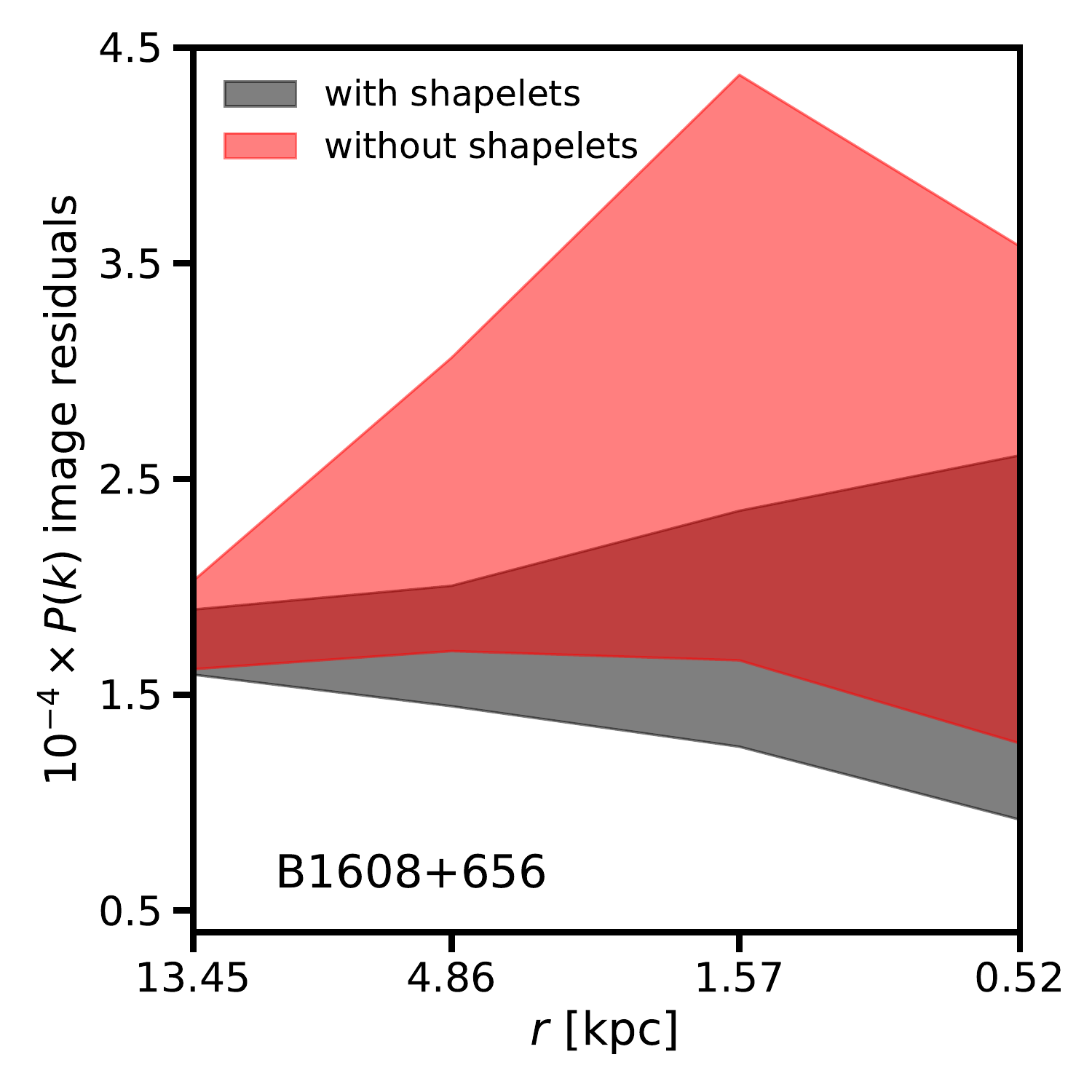}
		\includegraphics[clip,trim=0cm 0cm 0cm
		0cm,width=.45\textwidth,keepaspectratio]{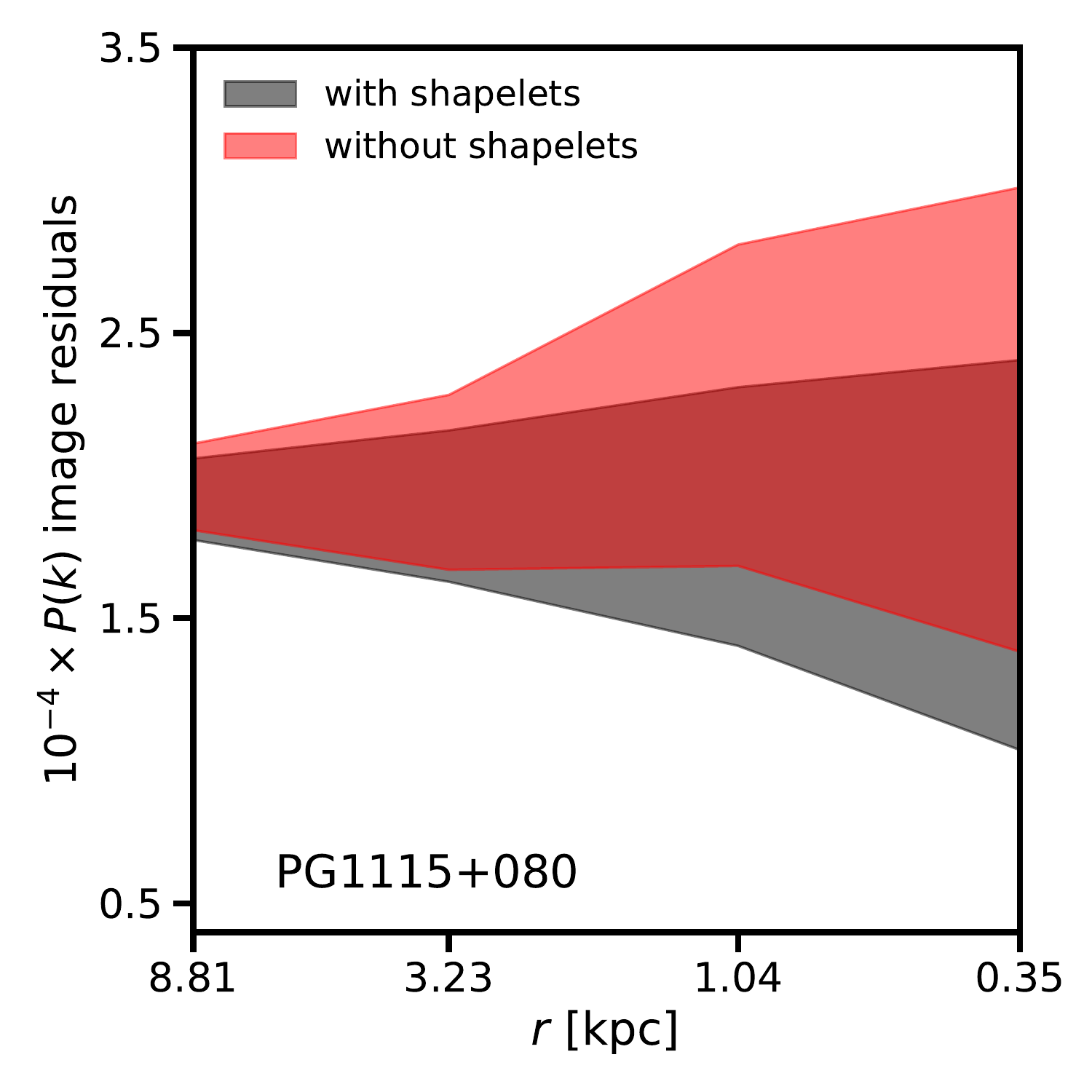}
		\caption{\label{fig:b1608pk} The stacked power spectrum of gravitational imaging residuals computed from 100 residual maps like those in Figure \ref{fig:b1608shapvsnoshap}. The manner in which substructure affects the power spectrum of imaging residuals on small scales depends on how the background source is modeled.}
	\end{figure*}

	In Figure \ref{fig:b1608pk}, we show the stacked power spectra of imaging residuals from 100 different realizations. The shaded regions represent show $68\%$ bootstrap confidence intervals around the median in each distance bin plotted on the x-axis. Without shapelets included in the background source, substructure leads to enhanced power on small scales, as expected. The addition of shapelets, which in this example act as a proxy for any additional source complexity that is typically required to model lensed images, removes the signal. 
	
	While the effects of very massive halos that produce direct hits on extended images evidently cannot be completely absorbed by the background source \citep[e.g.][]{Vegetti++10,Hezaveh++16b}, the collective effects of the many smaller halos that affect the convergence power spectrum can be partially absorbed by a reconfiguration of the lens and source model. The degree to which this occurs depends on the specifics of the source model, and on the quality of the imaging data. In principle, the fact that the source is multiply imaged breaks degeneracies between the source structure and imaging residuals \citep{CyrRacine++19}, but with finite resolution and signal to noise the effect persists. Higher-resolution images, targeting highly-magnified portions of extended images, and the identification of what scales are dominated by substructure perturbations rather than source morphology, may help deal with this systematic challenge. 
	
	
	\section{Joint posterior distributions of $H_0$ with lens macromodel parameters}	
	\label{sec:apph0macro}
	
\begin{figure*}
		\includegraphics[clip,trim=0.5cm 0.25cm 0.5cm
		0.5cm,width=.95\textwidth,keepaspectratio]{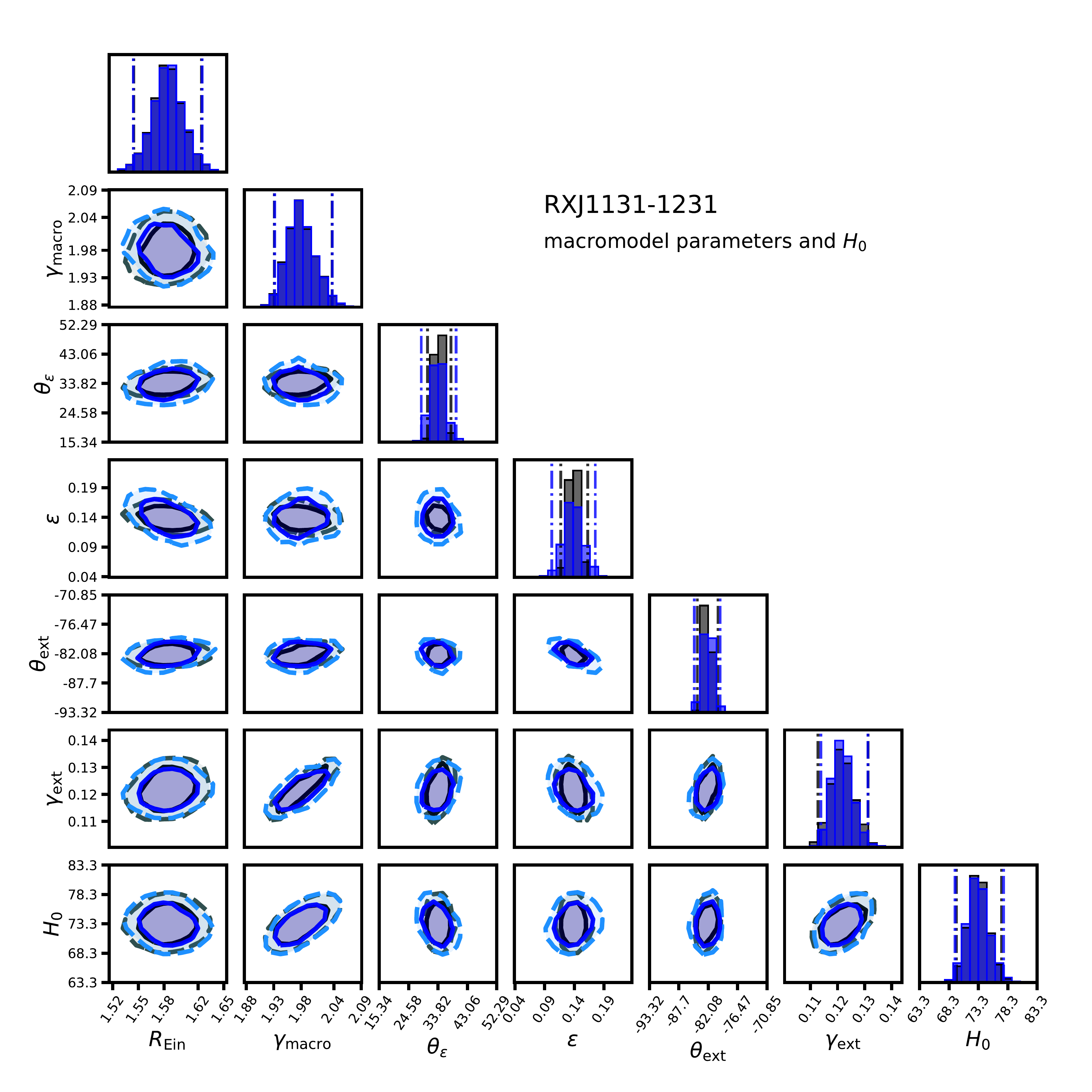}
		\caption{\label{fig:lens1131macroh0}  Joint distribution of selected macromodel parameters and the inferred Hubble constant for the control data set (black) and the substructure-perturbed dataset (blue). The parameters are (reading the x-axis labels from the left) the Einstein radius, the logarithmic slope of the main deflector, the ellipticity position angle, the ellipticity, the external shear position angle, the external shear strength, and the Hubble constant. Vertical lines show $95\%$ confidence intervals and the contours show $68 \%$ and $95\%$ confidence intervals. The distributions show are computed for the lens system RXJ1131-1231.}
	\end{figure*}

\begin{figure*}
		\includegraphics[clip,trim=0.5cm 0.25cm 0.5cm
		0.5cm,width=.95\textwidth,keepaspectratio]{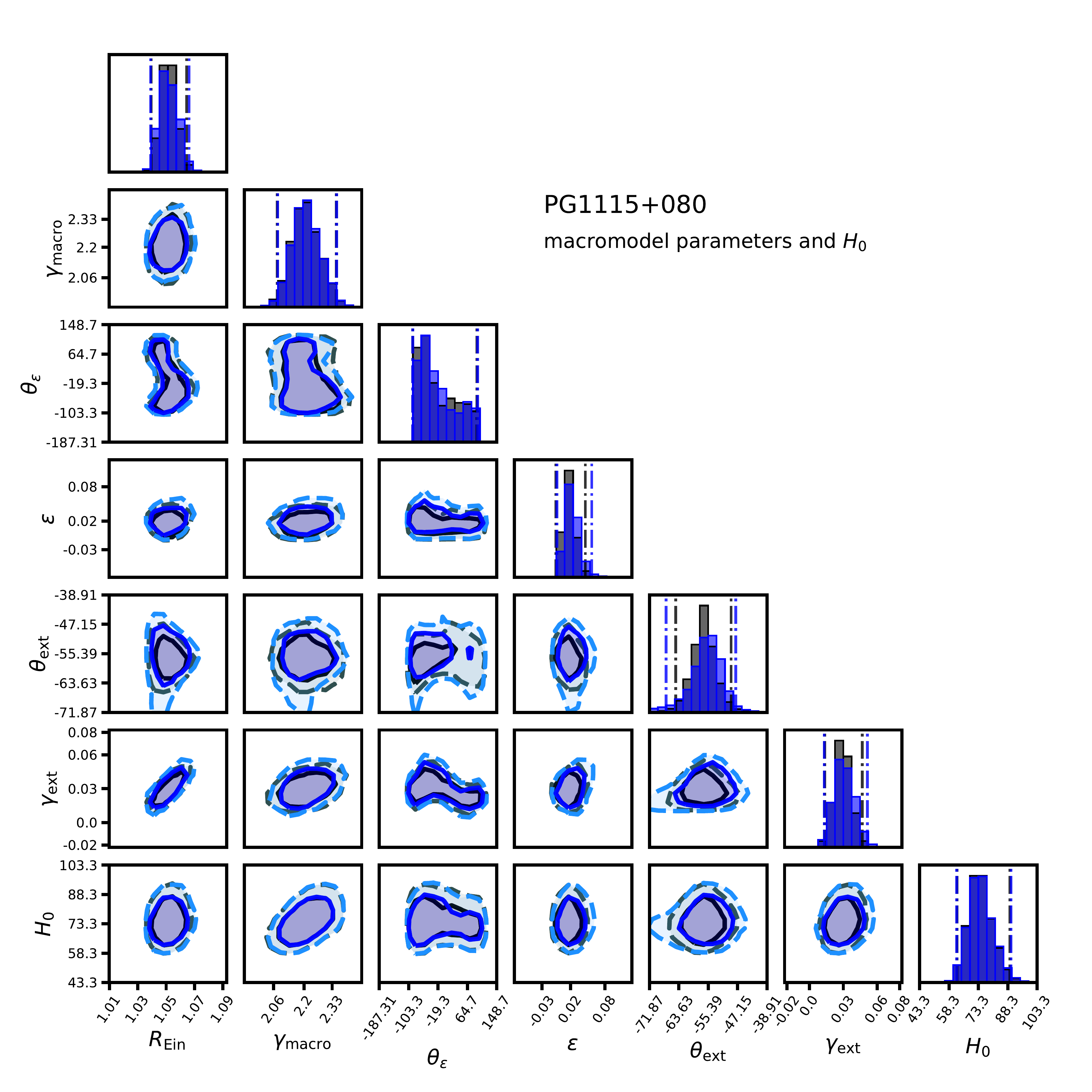}
		\caption{\label{fig:lens1115macroh0}  Joint distribution of selected macromodel parameters and the inferred Hubble constant for the lens system PG1115+080. See Figure \ref{fig:lens1131macroh0} for a description of the parameters.}
	\end{figure*}

\begin{figure*}
		\includegraphics[clip,trim=0.5cm 0.25cm 0.5cm
		0.5cm,width=.95\textwidth,keepaspectratio]{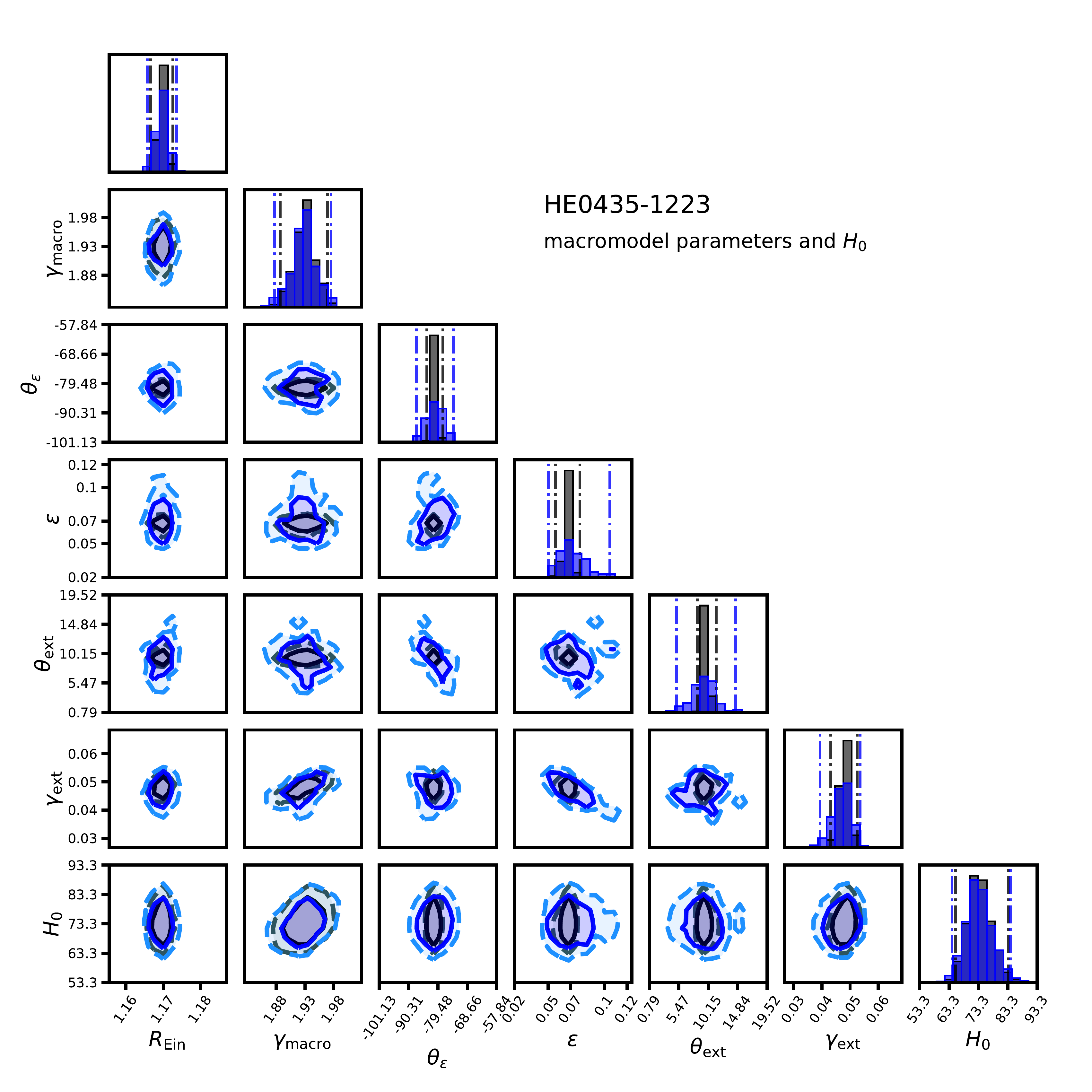}
		\caption{\label{fig:lens0435macroh0}   Joint distribution of selected macromodel parameters and the inferred Hubble constant for the lens system HE0435-1223. See Figure \ref{fig:lens1131macroh0} for a description of the parameters.}
	\end{figure*}

\begin{figure*}
		\includegraphics[clip,trim=0.5cm 0.25cm 0.5cm
		0.5cm,width=.95\textwidth,keepaspectratio]{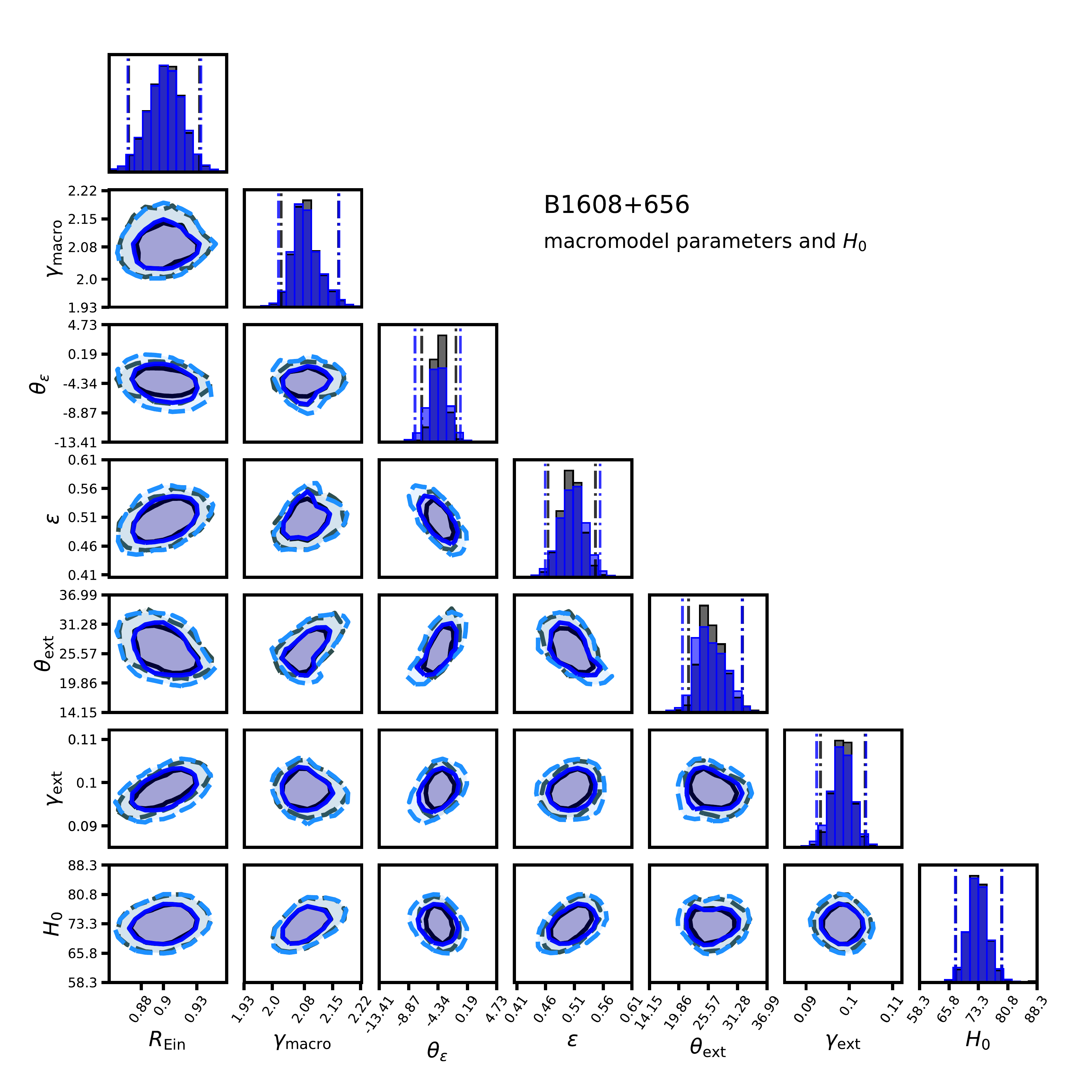}
		\caption{\label{fig:lens1608macroh0}  Joint distribution of selected macromodel parameters and the inferred Hubble constant for the lens system B1608+656. See Figure \ref{fig:lens1131macroh0} for a description of the parameters.}
	\end{figure*}

\begin{figure*}
		\includegraphics[clip,trim=0.5cm 0.25cm 0.5cm
		0.5cm,width=.95\textwidth,keepaspectratio]{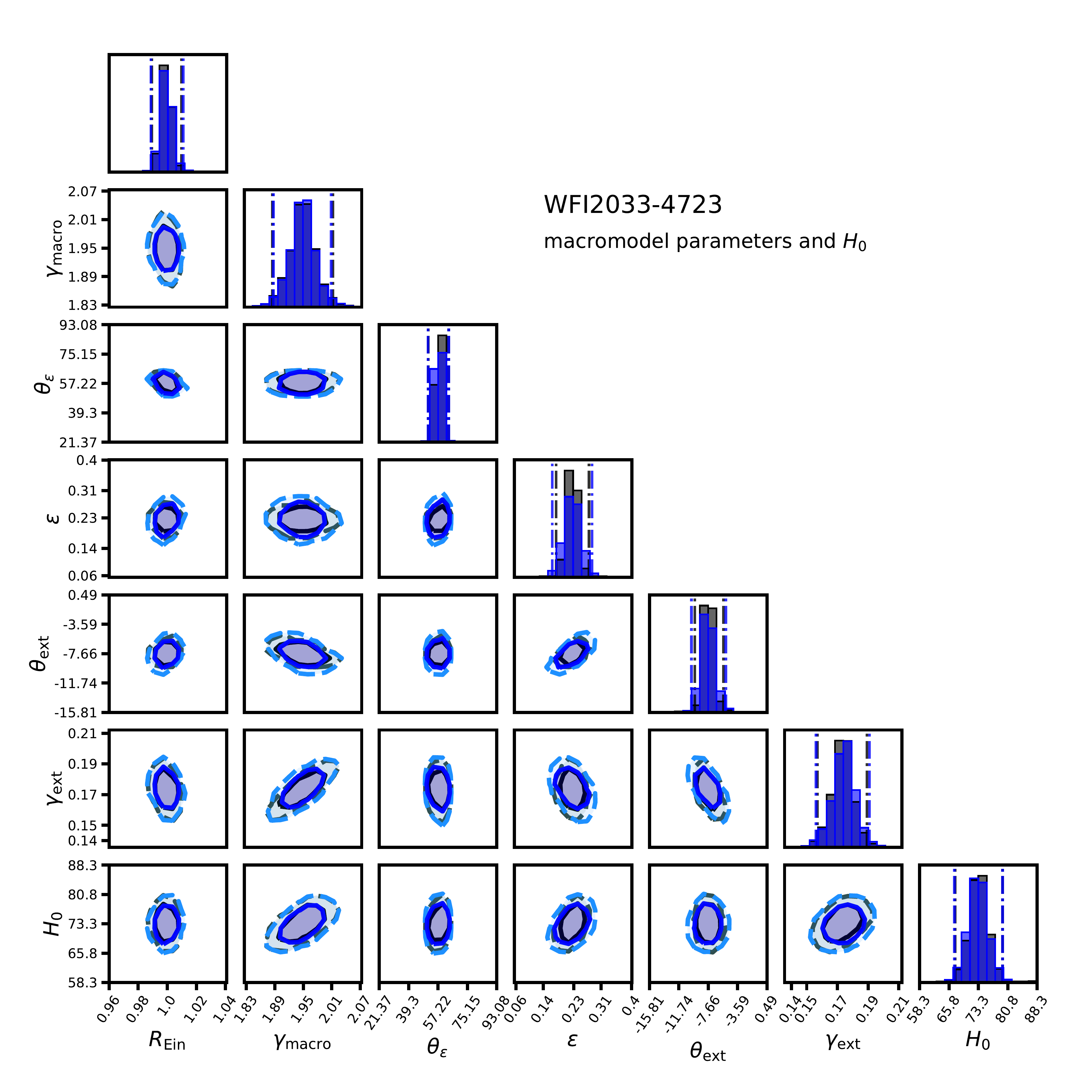}
		\caption{\label{fig:lens2033macroh0}  Joint distribution of selected macromodel parameters and the inferred Hubble constant for the lens system WFI2033-4723. See Figure \ref{fig:lens1131macroh0} for a description of the parameters.}
	\end{figure*}

\begin{figure*}
		\includegraphics[clip,trim=0.5cm 0.25cm 0.5cm
		0.5cm,width=.95\textwidth,keepaspectratio]{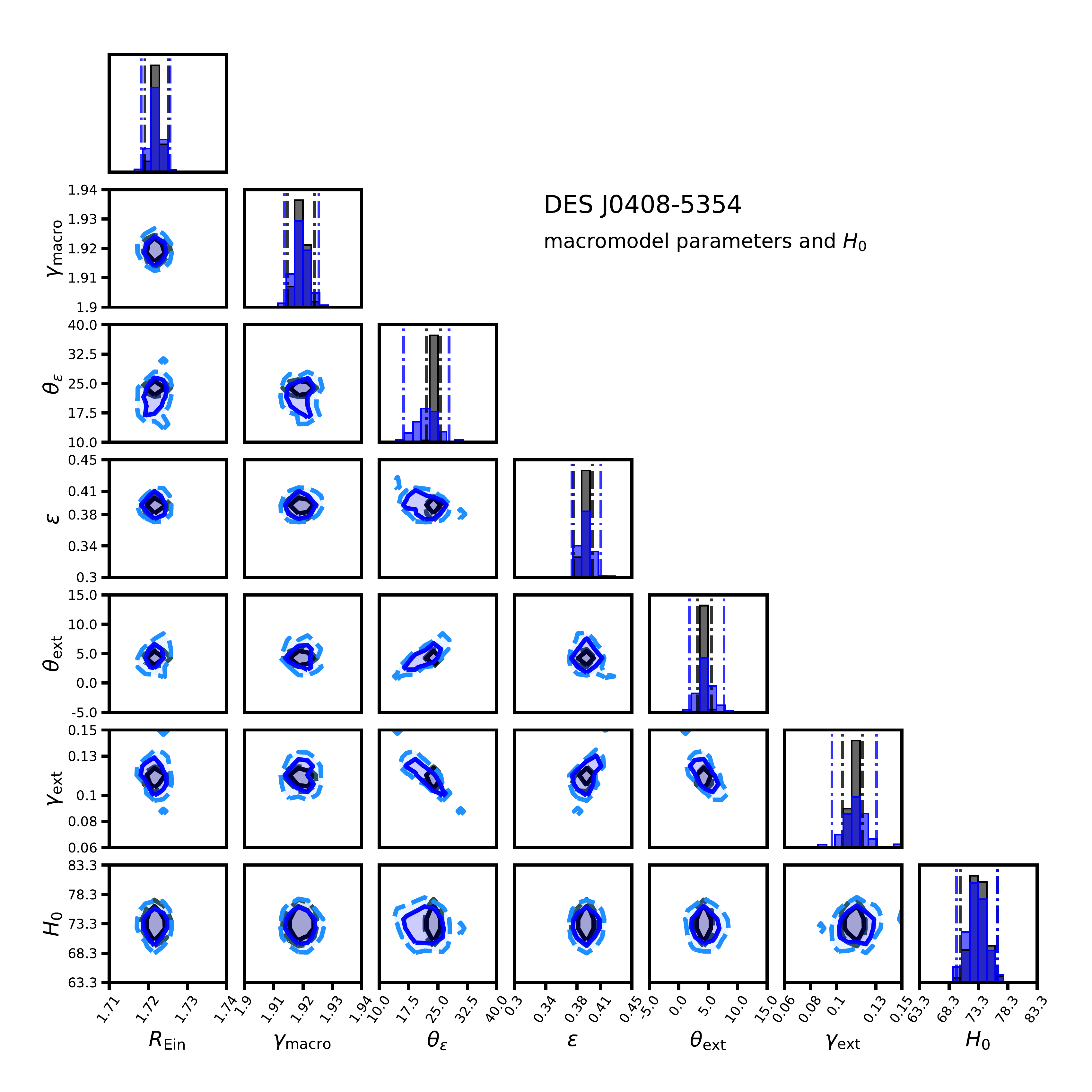}
		\caption{\label{fig:lens0408macroh0} Joint distribution of selected macromodel parameters and the inferred Hubble constant for the lens system DESJ0408-5354. See Figure \ref{fig:lens1131macroh0} for a description of the parameters. }
	\end{figure*}

	Figures \ref{fig:lens1131macroh0} through \ref{fig:lens0408macroh0} show the joint posterior distributions of macromodel parameters with the Hubble constant for the six lens systems analyzed in this work. 
	
	\section{Joint posterior distributions of time delays and flux ratios}
	\label{app:apptdelayfr}
	
	The initial motivation for considering the effects of substructure on time delays was to probe the nature of dark matter through these data. The thinking was that time delay `anomalies', or measured time delays that differ from the model-predicted time delays, encode the abundance and spatial distribution of dark substructure \citep[e.g.][]{KeetonMoustakas09,Cyr-Racine++16}. 
	
\begin{figure*}
		\includegraphics[clip,trim=0.5cm 0.25cm 0.5cm
		0.5cm,width=.95\textwidth,keepaspectratio]{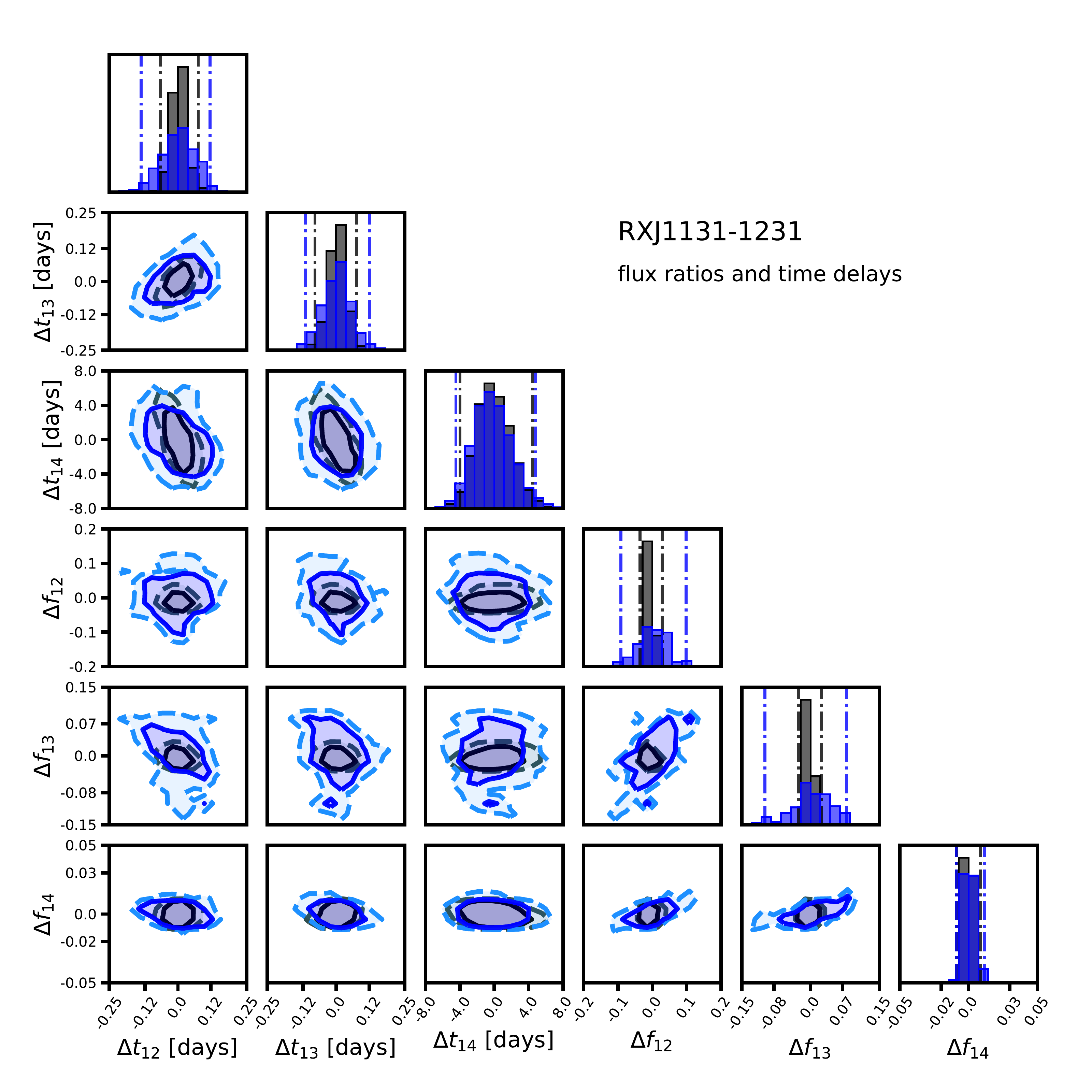}
		\caption{\label{fig:lens1131tdelayfr}  Joint distribution of flux ratio and time delay residuals (model minus data) computed for the lens system RXJ1131-1231.}
	\end{figure*}

\begin{figure*}
		\includegraphics[clip,trim=0.5cm 0.25cm 0.5cm
		0.5cm,width=.95\textwidth,keepaspectratio]{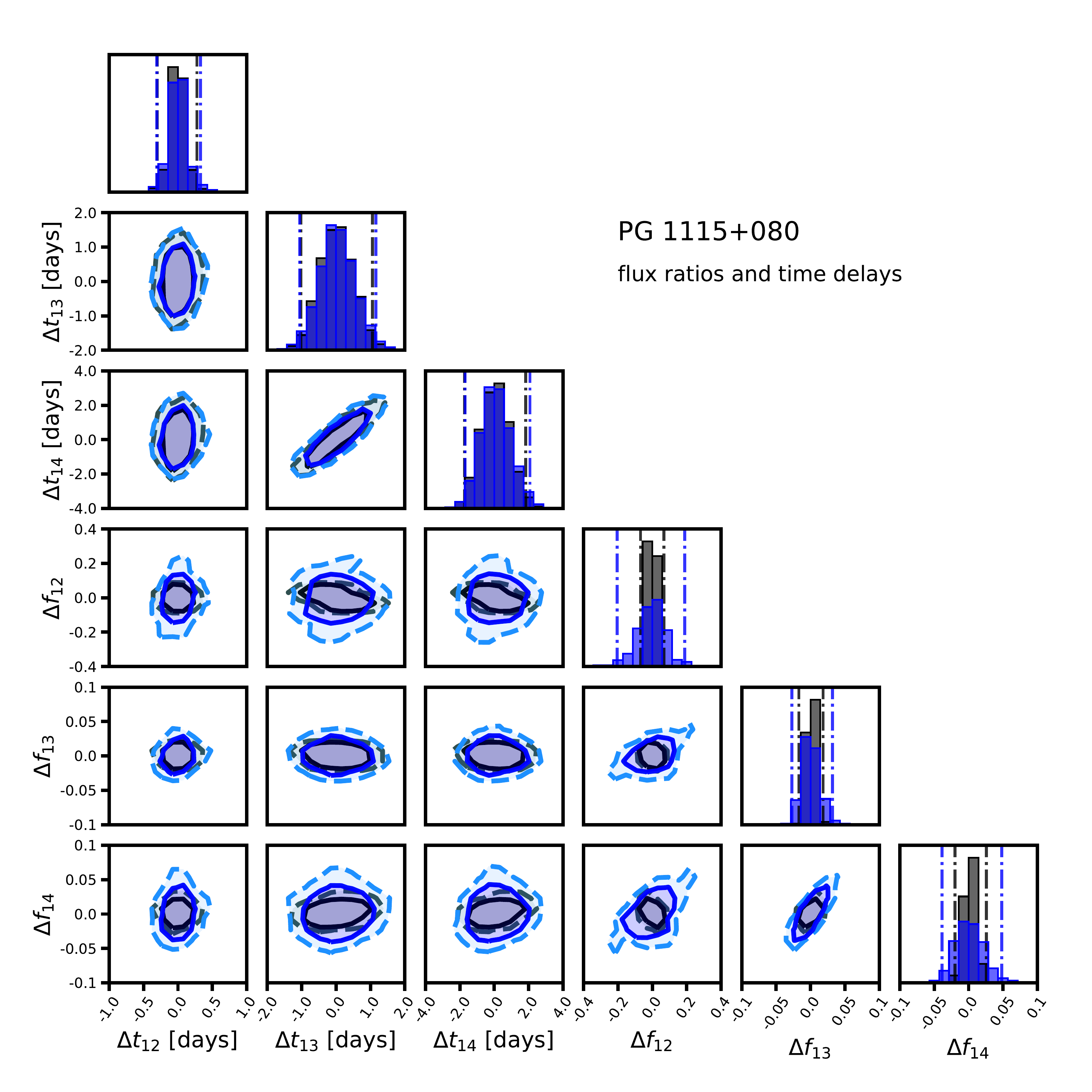}
		\caption{\label{fig:lens1115tdelayfr} Joint distribution of flux ratio and time delay residuals (model minus data) computed for the lens system PG1115+080. }
	\end{figure*}

\begin{figure*}
		\includegraphics[clip,trim=0.5cm 0.25cm 0.5cm
		0.5cm,width=.95\textwidth,keepaspectratio]{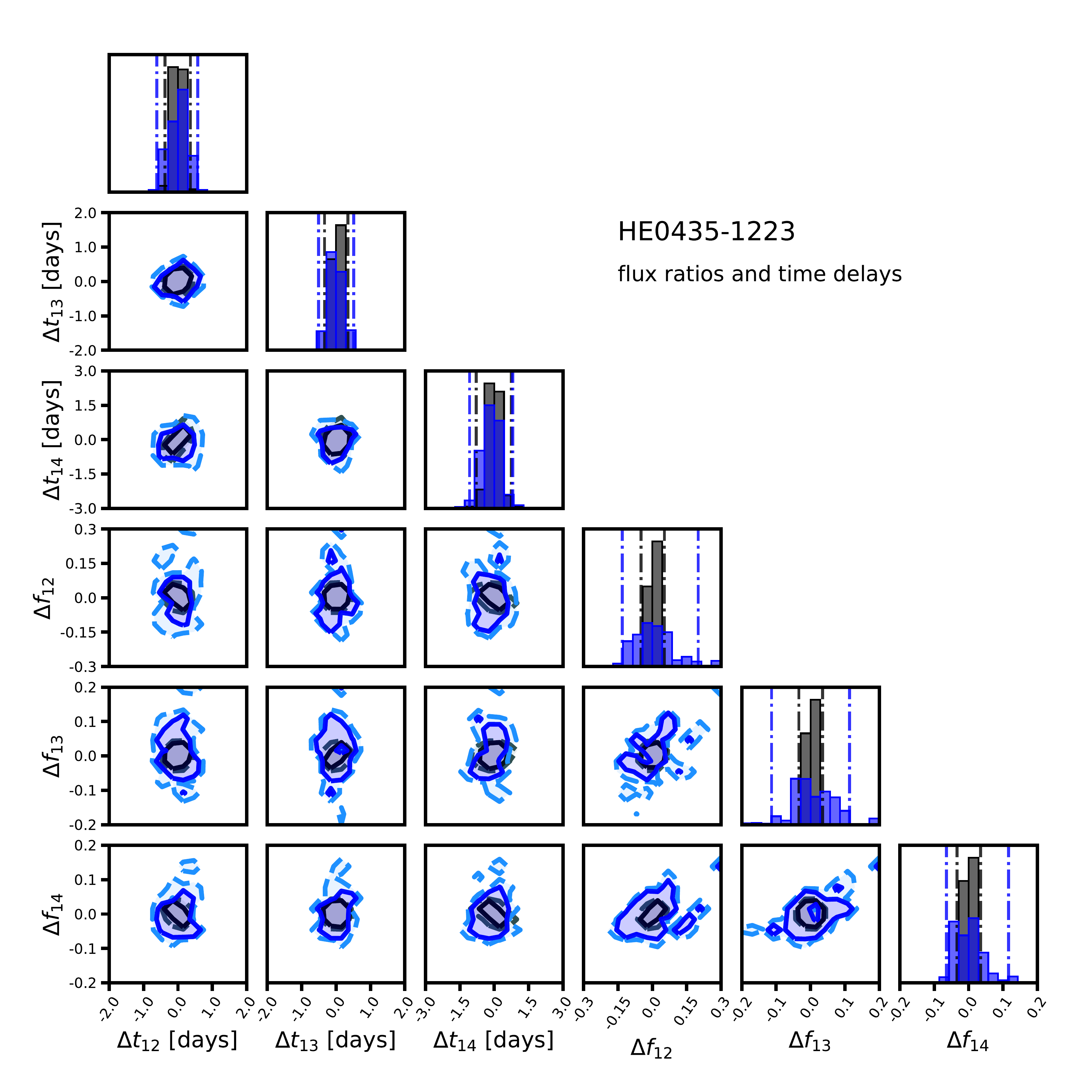}
		\caption{\label{fig:lens0435tdelayfr} Joint distribution of flux ratio and time delay residuals (model minus data) computed for the lens system HE0435-1223.}
	\end{figure*}

\begin{figure*}
		\includegraphics[clip,trim=0.5cm 0.25cm 0.5cm
		0.5cm,width=.95\textwidth,keepaspectratio]{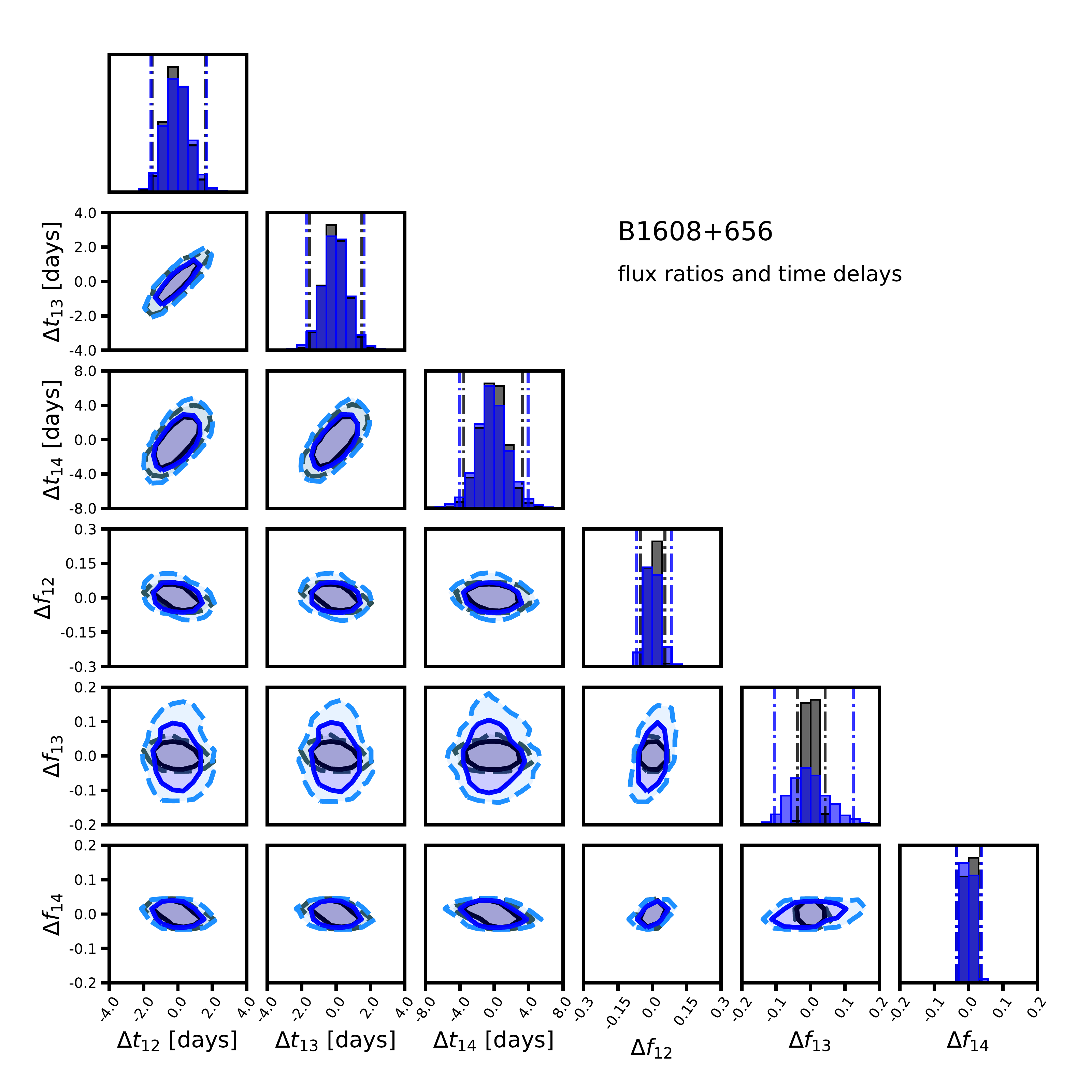}
		\caption{\label{fig:lens1608tdelayfr} Joint distribution of flux ratio and time delay residuals (model minus data) computed for the lens system B1608-656.}
	\end{figure*}

\begin{figure*}
		\includegraphics[clip,trim=0.5cm 0.25cm 0.5cm
		0.5cm,width=.95\textwidth,keepaspectratio]{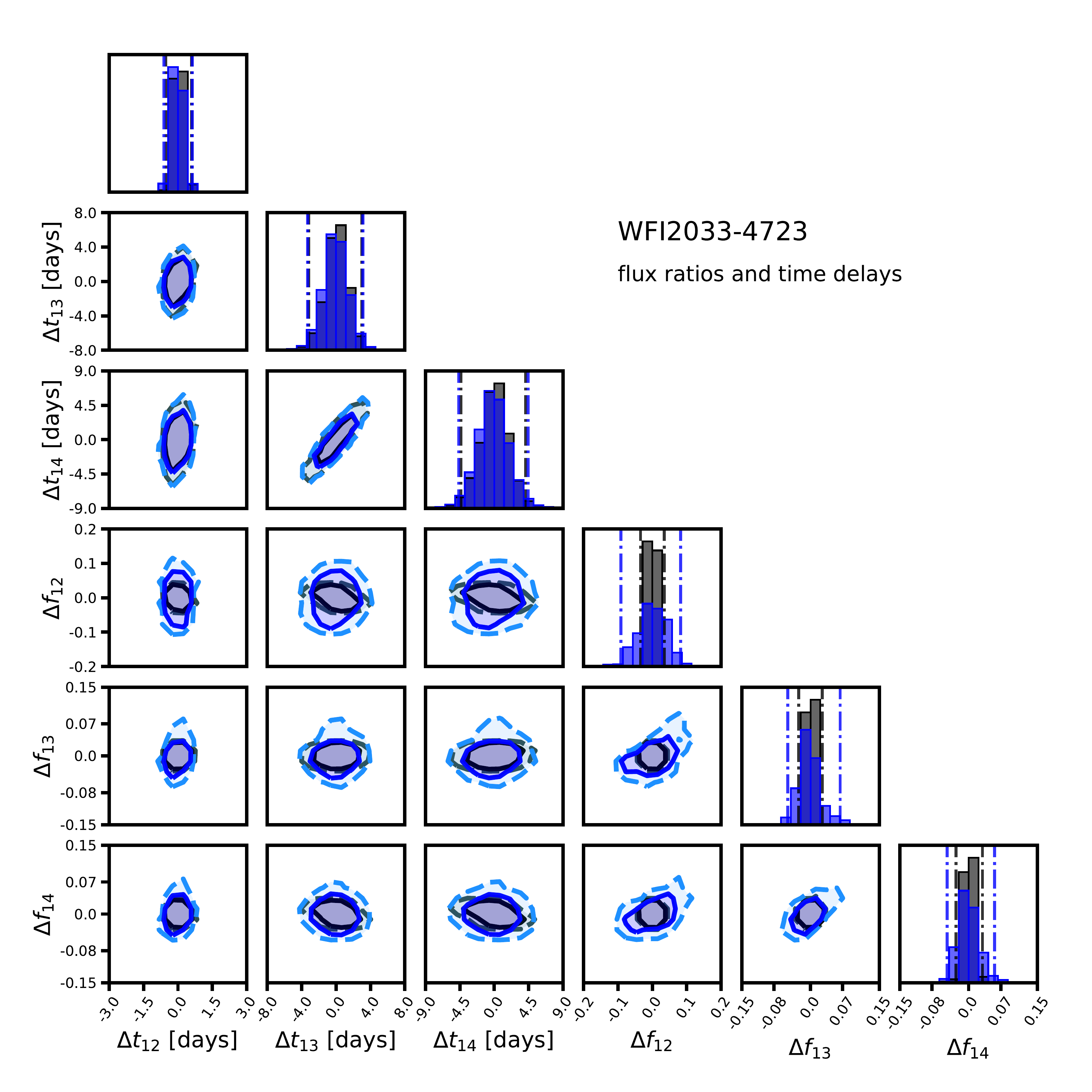}
		\caption{\label{fig:lens2033tdelayfr} Joint distribution of flux ratio and time delay residuals (model minus data) computed for the lens system WFI2033-4723. }
	\end{figure*}

\begin{figure*}
		\includegraphics[clip,trim=0.5cm 0.25cm 0.5cm
		0.5cm,width=.95\textwidth,keepaspectratio]{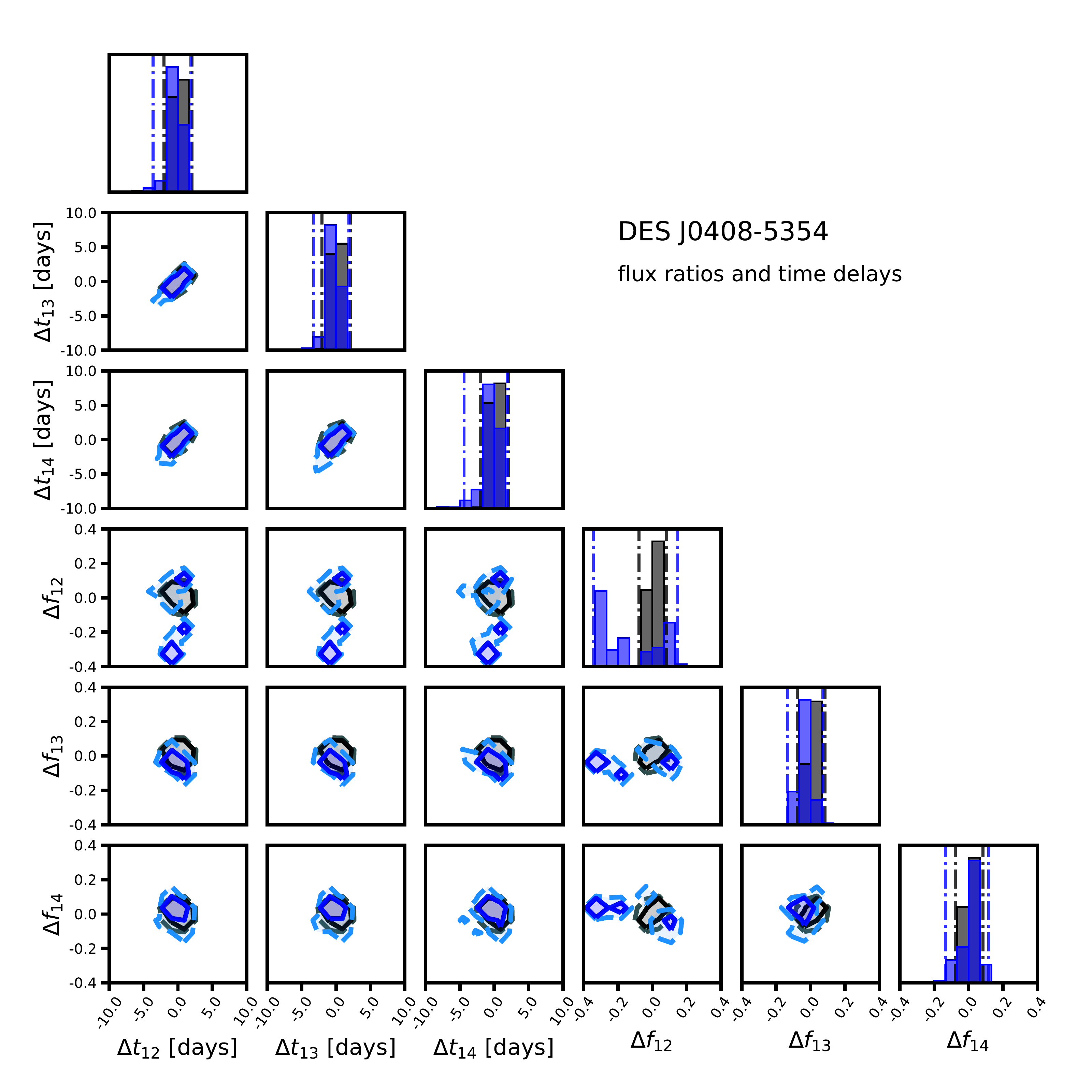}
		\caption{\label{fig:lens0408tdelayfr} Joint distribution of flux ratio and time delay residuals (model minus data) computed for the lens system DESJ0408-5354.}
	\end{figure*}

	The Shapiro delay associated with dark (sub)halos depends on the projected gravitational potential of halos, and is a long-range interaction. This could, in principle, complement the information encoded by image flux ratios, which are highly localized probes of dark matter structure and provide an avenue to probe small-scale structure in strong lenses \citep{DalalKochanek02,Nierenberg++14,Hsueh++19,Gilman++20a,Gilman++20b}. Figures \ref{fig:lens1131tdelayfr} through \ref{fig:lens0408tdelayfr} show the joint distributions of the residual time delays and residual flux ratios for the six lens systems analyzed. The flux ratios are computed with a background source parameterized as a Gaussian with a FWHM of 25 pc. 
	
\end{document}